\documentclass{article}

\usepackage{amssymb}
\usepackage{amsmath}

\newtheorem{theorem}{Theorem}

\newtheorem{axiom}[theorem]{Axiom}

\newtheorem{conjecture}[theorem]{Conjecture}
\newtheorem{corollary}[theorem]{Corollary}

\newtheorem{definition}[theorem]{Definition}
\newtheorem{example}[theorem]{Example}
\newtheorem{exercise}[theorem]{Exercise}
\newtheorem{lemma}[theorem]{Lemma}

\newtheorem{proposition}[theorem]{Proposition}
\newtheorem{remark}[theorem]{Remark}

\newenvironment{proof}[1][Proof]{\noindent\textbf{#1.} }{\ \rule{0.5em}{0.5em}}

\makeatletter

\newcount\@hour\newcount\@minute\chardef\@x10\chardef\@xv60
\def\tcitime{
\def\@time{%
  \@minute\time\@hour\@minute\divide\@hour\@xv
  \ifnum\@hour<\@x 0\fi\the\@hour:%
  \multiply\@hour\@xv\advance\@minute-\@hour
  \ifnum\@minute<\@x 0\fi\the\@minute
  }}%

\def\x@hyperref#1#2#3{%
   \catcode`\~ = 12
   \catcode`\$ = 12
   \catcode`\_ = 12
   \catcode`\# = 12
   \catcode`\& = 12
   \y@hyperref{#1}{#2}{#3}%
}

\def\y@hyperref#1#2#3#4{%
   #2\ref{#4}#3
   \catcode`\~ = 13
   \catcode`\$ = 3
   \catcode`\_ = 8
   \catcode`\# = 6
   \catcode`\& = 4
}

\@ifundefined{hyperref}{\let\hyperref\x@hyperref}{}
\@ifundefined{msihyperref}{\let\msihyperref\x@hyperref}{}

\@ifundefined{qExtProgCall}{\def\qExtProgCall#1#2#3#4#5#6{\relax}}{}
\def\QCTOpt[#1]#2{%
  \def\QCTOptB{#1}
  \def\QCTOptA{#2}
}
\def\QCTNOpt#1{%
  \def\QCTOptA{#1}
  \let\QCTOptB\empty
}
\def\Qct{%
  \@ifnextchar[{%
    \QCTOpt}{\QCTNOpt}
}
\def\QCBOpt[#1]#2{%
  \def\QCBOptB{#1}%
  \def\QCBOptA{#2}%
}
\def\QCBNOpt#1{%
  \def\QCBOptA{#1}%
  \let\QCBOptB\empty
}
\def\Qcb{%
  \@ifnextchar[{%
    \QCBOpt}{\QCBNOpt}%
}
\def\PrepCapArgs{%
  \ifx\QCBOptA\empty
    \ifx\QCTOptA\empty
      {}%
    \else
      \ifx\QCTOptB\empty
        {\QCTOptA}%
      \else
        [\QCTOptB]{\QCTOptA}%
      \fi
    \fi
  \else
    \ifx\QCBOptA\empty
      {}%
    \else
      \ifx\QCBOptB\empty
        {\QCBOptA}%
      \else
        [\QCBOptB]{\QCBOptA}%
      \fi
    \fi
  \fi
}
\newcount\GRAPHICSTYPE
\GRAPHICSTYPE=\z@
\def\GRAPHICSPS#1{%
 \ifcase\GRAPHICSTYPE
   \special{ps: #1}%
 \or
   \special{language "PS", include "#1"}%
 \fi
}%
\def\graffile#1#2#3#4{%
    \bgroup
	   \@inlabelfalse
       \leavevmode
       \@ifundefined{bbl@deactivate}{\def~{\string~}}{\activesoff}%
        \raise -#4 \BOXTHEFRAME{%
           \hbox to #2{\raise #3\hbox to #2{\null #1\hfil}}}%
    \egroup
}%
\def\draftbox#1#2#3#4{%
 \leavevmode\raise -#4 \hbox{%
  \frame{\rlap{\protect\tiny #1}\hbox to #2%
   {\vrule height#3 width\z@ depth\z@\hfil}%
  }%
 }%
}%
\newcount\@msidraft
\@msidraft=\z@
\let\nographics=\@msidraft
\newif\ifwasdraft
\wasdraftfalse

\def\GRAPHIC#1#2#3#4#5{%
   \ifnum\@msidraft=\@ne\draftbox{#2}{#3}{#4}{#5}%
   \else\graffile{#1}{#3}{#4}{#5}%
   \fi
}
\def\addtoLaTeXparams#1{%
    \edef\LaTeXparams{\LaTeXparams #1}}%
\newif\ifBoxFrame \BoxFramefalse
\newif\ifOverFrame \OverFramefalse
\newif\ifUnderFrame \UnderFramefalse

\def\BOXTHEFRAME#1{%
   \hbox{%
      \ifBoxFrame
         \frame{#1}%
      \else
         {#1}%
      \fi
   }%
}

\def\doFRAMEparams#1{\BoxFramefalse\OverFramefalse\UnderFramefalse\readFRAMEparams#1\end}%
\def\readFRAMEparams#1{%
 \ifx#1\end%
  \let\next=\relax
  \else
  \ifx#1i\dispkind=\z@\fi
  \ifx#1d\dispkind=\@ne\fi
  \ifx#1f\dispkind=\tw@\fi
  \ifx#1t\addtoLaTeXparams{t}\fi
  \ifx#1b\addtoLaTeXparams{b}\fi
  \ifx#1p\addtoLaTeXparams{p}\fi
  \ifx#1h\addtoLaTeXparams{h}\fi
  \ifx#1X\BoxFrametrue\fi
  \ifx#1O\OverFrametrue\fi
  \ifx#1U\UnderFrametrue\fi
  \ifx#1w
    \ifnum\@msidraft=1\wasdrafttrue\else\wasdraftfalse\fi
    \@msidraft=\@ne
  \fi
  \let\next=\readFRAMEparams
  \fi
 \next
 }%
\def\IFRAME#1#2#3#4#5#6{%
      \bgroup
      \let\QCTOptA\empty
      \let\QCTOptB\empty
      \let\QCBOptA\empty
      \let\QCBOptB\empty
      #6%
      \parindent=0pt
      \leftskip=0pt
      \rightskip=0pt
      \setbox0=\hbox{\QCBOptA}%
      \@tempdima=#1\relax
      \ifOverFrame
          \typeout{This is not implemented yet}%
          \show\HELP
      \else
         \ifdim\wd0>\@tempdima
            \advance\@tempdima by \@tempdima
            \ifdim\wd0 >\@tempdima
               \setbox1 =\vbox{%
                  \unskip\hbox to \@tempdima{\hfill\GRAPHIC{#5}{#4}{#1}{#2}{#3}\hfill}%
                  \unskip\hbox to \@tempdima{\parbox[b]{\@tempdima}{\QCBOptA}}%
               }%
               \wd1=\@tempdima
            \else
               \textwidth=\wd0
               \setbox1 =\vbox{%
                 \noindent\hbox to \wd0{\hfill\GRAPHIC{#5}{#4}{#1}{#2}{#3}\hfill}\\%
                 \noindent\hbox{\QCBOptA}%
               }%
               \wd1=\wd0
            \fi
         \else
            \ifdim\wd0>0pt
              \hsize=\@tempdima
              \setbox1=\vbox{%
                \unskip\GRAPHIC{#5}{#4}{#1}{#2}{0pt}%
                \break
                \unskip\hbox to \@tempdima{\hfill \QCBOptA\hfill}%
              }%
              \wd1=\@tempdima
           \else
              \hsize=\@tempdima
              \setbox1=\vbox{%
                \unskip\GRAPHIC{#5}{#4}{#1}{#2}{0pt}%
              }%
              \wd1=\@tempdima
           \fi
         \fi
         \@tempdimb=\ht1
         \advance\@tempdimb by -#2
         \advance\@tempdimb by #3
         \leavevmode
         \raise -\@tempdimb \hbox{\box1}%
      \fi
      \egroup%
}%
\def\DFRAME#1#2#3#4#5{%
  \hfil\break
  \bgroup
     \leftskip\@flushglue
	 \rightskip\@flushglue
	 \parindent\z@
	 \parfillskip\z@skip
     \let\QCTOptA\empty
     \let\QCTOptB\empty
     \let\QCBOptA\empty
     \let\QCBOptB\empty
	 \vbox\bgroup
        \ifOverFrame 
           #5\QCTOptA\par
        \fi
        \GRAPHIC{#4}{#3}{#1}{#2}{\z@}%
        \ifUnderFrame 
           \break#5\QCBOptA
        \fi
	 \egroup
   \egroup
   \break
}%
\def\FFRAME#1#2#3#4#5#6#7{%
  \@ifundefined{floatstyle}
    {
     \begin{figure}[#1]%
    }
    {
	 \ifx#1h
      \begin{figure}[H]%
	 \else
      \begin{figure}[#1]%
	 \fi
	}
  \let\QCTOptA\empty
  \let\QCTOptB\empty
  \let\QCBOptA\empty
  \let\QCBOptB\empty
  \ifOverFrame
    #4
    \ifx\QCTOptA\empty
    \else
      \ifx\QCTOptB\empty
        \caption{\QCTOptA}%
      \else
        \caption[\QCTOptB]{\QCTOptA}%
      \fi
    \fi
    \ifUnderFrame\else
      \label{#5}%
    \fi
  \else
    \UnderFrametrue%
  \fi
  \begin{center}\GRAPHIC{#7}{#6}{#2}{#3}{\z@}\end{center}%
  \ifUnderFrame
    #4
    \ifx\QCBOptA\empty
      \caption{}%
    \else
      \ifx\QCBOptB\empty
        \caption{\QCBOptA}%
      \else
        \caption[\QCBOptB]{\QCBOptA}%
      \fi
    \fi
    \label{#5}%
  \fi
  \end{figure}%
 }%
\newcount\dispkind%

\def\makeactives{
  \catcode`\"=\active
  \catcode`\;=\active
  \catcode`\:=\active
  \catcode`\'=\active
  \catcode`\~=\active
}
\bgroup
   \makeactives
   \gdef\activesoff{%
      \def"{\string"}
      \def;{\string;}
      \def:{\string:}
      \def'{\string'}
      \def~{\string~}
    }
\egroup

\def\FRAME#1#2#3#4#5#6#7#8{%
 \bgroup
 \ifnum\@msidraft=\@ne
   \wasdrafttrue
 \else
   \wasdraftfalse%
 \fi
 \def\LaTeXparams{}%
 \dispkind=\z@
 \def\LaTeXparams{}%
 \doFRAMEparams{#1}%
 \ifnum\dispkind=\z@\IFRAME{#2}{#3}{#4}{#7}{#8}{#5}\else
  \ifnum\dispkind=\@ne\DFRAME{#2}{#3}{#7}{#8}{#5}\else
   \ifnum\dispkind=\tw@
    \edef\@tempa{\noexpand\FFRAME{\LaTeXparams}}%
    \@tempa{#2}{#3}{#5}{#6}{#7}{#8}%
    \fi
   \fi
  \fi
  \ifwasdraft\@msidraft=1\else\@msidraft=0\fi{}%
  \egroup
 }%
\def\TEXUX#1{"texux"}

\long\def\QQQ#1#2{%
     \long\expandafter\def\csname#1\endcsname{#2}}%
\@ifundefined{QTP}{\def\QTP#1{}}{}
\@ifundefined{QEXCLUDE}{\def\QEXCLUDE#1{}}{}
\@ifundefined{Qlb}{}{}
\@ifundefined{Qlt}{}{}
\long\def\QQA#1#2{}%
\def\QTR#1#2{{\csname#1\endcsname #2}}
\def\EXPAND#1[#2]#3{}%
\def\NOEXPAND#1[#2]#3{}%
\def\LaTeXparent#1{}%
\def\ChildStyles#1{}%
\def\ChildDefaults#1{}%
\def\QTagDef#1#2#3{}%

\@ifundefined{correctchoice}{}{}
\@ifundefined{HTML}{\def\HTML#1{\relax}}{}
\@ifundefined{TCIIcon}{\def\TCIIcon#1#2#3#4{\relax}}{}
\if@compatibility
\else
  \providecommand{\UNICODE}[2][]{\protect\rule{.1in}{.1in}}
  \providecommand{\U}[1]{\protect\rule{.1in}{.1in}}
  
\fi

\@ifundefined{lambdabar}{
      
   }{}

\@ifundefined{StyleEditBeginDoc}{}{}
\def\QQfnmark#1{\footnotemark}

\@ifundefined{TCIMAKEINDEX}{}{\makeindex}%
\@ifundefined{abstract}{%
 \def\abstract{%
  \if@twocolumn
   \section*{Abstract (Not appropriate in this style!)}%
   \else \small 
   \begin{center}{\bf Abstract\vspace{-.5em}\vspace{\z@}}\end{center}%
   \quotation 
   \fi
  }%
 }{%
 }%
\@ifundefined{endabstract}{\def\endabstract
  {\if@twocolumn\else\endquotation\fi}}{}%
\@ifundefined{maketitle}{\def\maketitle#1{}}{}%
\@ifundefined{affiliation}{\def\affiliation#1{}}{}%
\@ifundefined{proof}{}{}%
\@ifundefined{endproof}{}{}%
\@ifundefined{newfield}{\def\newfield#1#2{}}{}%
\@ifundefined{chapter}{\def\chapter#1{\par(Chapter head:)#1\par }%
 \newcount\c@chapter}{}%
\@ifundefined{part}{\def\part#1{\par(Part head:)#1\par }}{}%
\@ifundefined{section}{\def\section#1{\par(Section head:)#1\par }}{}%
\@ifundefined{subsection}{\def\subsection#1%
 {\par(Subsection head:)#1\par }}{}%
\@ifundefined{subsubsection}{\def\subsubsection#1%
 {\par(Subsubsection head:)#1\par }}{}%
\@ifundefined{paragraph}{\def\paragraph#1%
 {\par(Subsubsubsection head:)#1\par }}{}%
\@ifundefined{subparagraph}{\def\subparagraph#1%
 {\par(Subsubsubsubsection head:)#1\par }}{}%
\@ifundefined{therefore}{}{}%
\@ifundefined{backepsilon}{}{}%
\@ifundefined{yen}{}{}%
\@ifundefined{registered}{%
   \def\registered{\relax\ifmmode{}\r@gistered
                    \else$\m@th\r@gistered$\fi}%
 \def\r@gistered{^{\ooalign
  {\hfil\raise.07ex\hbox{$\scriptstyle\rm\text{R}$}\hfil\crcr
  \mathhexbox20D}}}}{}%
\@ifundefined{Eth}{}{}%
\@ifundefined{eth}{}{}%
\@ifundefined{Thorn}{}{}%
\@ifundefined{thorn}{}{}%
\@ifundefined{degree}{}{}%
\newdimen\theight
\@ifundefined{Column}{\def\Column{%
 \vadjust{\setbox\z@=\hbox{\scriptsize\quad\quad tcol}%
  \theight=\ht\z@\advance\theight by \dp\z@\advance\theight by \lineskip
  \kern -\theight \vbox to \theight{%
   \rightline{\rlap{\box\z@}}%
   \vss
   }%
  }%
 }}{}%
\@ifundefined{qed}{\def\qed{%
 \ifhmode\unskip\nobreak\fi\ifmmode\ifinner\else\hskip5\p@\fi\fi
 \hbox{\hskip5\p@\vrule width4\p@ height6\p@ depth1.5\p@\hskip\p@}%
 }}{}%
\@ifundefined{cents}{}{}%
\@ifundefined{tciLaplace}{}{}%
\@ifundefined{tciFourier}{}{}%
\@ifundefined{textcurrency}{}{}%
\@ifundefined{texteuro}{}{}%
\@ifundefined{textfranc}{}{}%
\@ifundefined{textlira}{}{}%
\@ifundefined{textpeseta}{}{}%
\@ifundefined{miss}{\def\miss{\hbox{\vrule height2\p@ width 2\p@ depth\z@}}}{}%
\@ifundefined{vvert}{}{}
\@ifundefined{tcol}{\def\tcol#1{{\baselineskip=6\p@ \vcenter{#1}} \Column}}{}%
\@ifundefined{dB}{}{}
\@ifundefined{mB}{}{}
\@ifundefined{nB}{}{}
\def\newfmtname{LaTeX2e}
\ifx\fmtname\newfmtname
  \DeclareOldFontCommand{\rm}{\normalfont\rmfamily}{\mathrm}
  \DeclareOldFontCommand{\sf}{\normalfont\sffamily}{\mathsf}
  \DeclareOldFontCommand{\tt}{\normalfont\ttfamily}{\mathtt}
  \DeclareOldFontCommand{\bf}{\normalfont\bfseries}{\mathbf}
  \DeclareOldFontCommand{\it}{\normalfont\itshape}{\mathit}
  \DeclareOldFontCommand{\sl}{\normalfont\slshape}{\@nomath\sl}
  \DeclareOldFontCommand{\sc}{\normalfont\scshape}{\@nomath\sc}
\fi

\def\alpha{{\Greekmath 010B}}%
\def\beta{{\Greekmath 010C}}%
\def\gamma{{\Greekmath 010D}}%
\def\delta{{\Greekmath 010E}}%
\def\epsilon{{\Greekmath 010F}}%
\def\zeta{{\Greekmath 0110}}%
\def\eta{{\Greekmath 0111}}%
\def\theta{{\Greekmath 0112}}%
\def\iota{{\Greekmath 0113}}%
\def\kappa{{\Greekmath 0114}}%
\def\lambda{{\Greekmath 0115}}%
\def\mu{{\Greekmath 0116}}%
\def\nu{{\Greekmath 0117}}%
\def\xi{{\Greekmath 0118}}%
\def\pi{{\Greekmath 0119}}%
\def\rho{{\Greekmath 011A}}%
\def\sigma{{\Greekmath 011B}}%
\def\tau{{\Greekmath 011C}}%
\def\upsilon{{\Greekmath 011D}}%
\def\phi{{\Greekmath 011E}}%
\def\chi{{\Greekmath 011F}}%
\def\psi{{\Greekmath 0120}}%
\def\omega{{\Greekmath 0121}}%
\def\varepsilon{{\Greekmath 0122}}%
\def\vartheta{{\Greekmath 0123}}%
\def\varpi{{\Greekmath 0124}}%
\def\varrho{{\Greekmath 0125}}%
\def\varsigma{{\Greekmath 0126}}%
\def\varphi{{\Greekmath 0127}}%

\def\nabla{{\Greekmath 0272}}
\def\FindBoldGroup{%
   {\setbox0=\hbox{$\mathbf{x\global\edef\theboldgroup{\the\mathgroup}}$}}%
}

\def\Greekmath#1#2#3#4{%
    \if@compatibility
        \ifnum\mathgroup=\symbold
           \mathchoice{\mbox{\boldmath$\displaystyle\mathchar"#1#2#3#4$}}%
                      {\mbox{\boldmath$\textstyle\mathchar"#1#2#3#4$}}%
                      {\mbox{\boldmath$\scriptstyle\mathchar"#1#2#3#4$}}%
                      {\mbox{\boldmath$\scriptscriptstyle\mathchar"#1#2#3#4$}}%
        \else
           \mathchar"#1#2#3#4%
        \fi 
    \else 
        \FindBoldGroup
        \ifnum\mathgroup=\theboldgroup 
           \mathchoice{\mbox{\boldmath$\displaystyle\mathchar"#1#2#3#4$}}%
                      {\mbox{\boldmath$\textstyle\mathchar"#1#2#3#4$}}%
                      {\mbox{\boldmath$\scriptstyle\mathchar"#1#2#3#4$}}%
                      {\mbox{\boldmath$\scriptscriptstyle\mathchar"#1#2#3#4$}}%
        \else
           \mathchar"#1#2#3#4%
        \fi     	    
	  \fi}

\newif\ifGreekBold  \GreekBoldfalse
\let\SAVEPBF=\pbf
\def\pbf{\GreekBoldtrue\SAVEPBF}%

\@ifundefined{theorem}{}{}
\@ifundefined{lemma}{}{}
\@ifundefined{corollary}{}{}
\@ifundefined{conjecture}{}{}
\@ifundefined{proposition}{}{}
\@ifundefined{axiom}{}{}
\@ifundefined{remark}{}{}
\@ifundefined{example}{}{}
\@ifundefined{exercise}{}{}
\@ifundefined{definition}{}{}

\@ifundefined{mathletters}{%
  \newcounter{equationnumber}  
  \def\mathletters{%
     \addtocounter{equation}{1}
     \edef\@currentlabel{\theequation}%
     \setcounter{equationnumber}{\c@equation}
     \setcounter{equation}{0}%
     \edef\theequation{\@currentlabel\noexpand\alph{equation}}%
  }
  
}{}

\@ifundefined{BibTeX}{%
    \def\BibTeX{{\rm B\kern-.05em{\sc i\kern-.025em b}\kern-.08em
                 T\kern-.1667em\lower.7ex\hbox{E}\kern-.125emX}}}{}%
\@ifundefined{AmS}%
    {\def\AmS{{\protect\usefont{OMS}{cmsy}{m}{n}%
                A\kern-.1667em\lower.5ex\hbox{M}\kern-.125emS}}}{}%
\@ifundefined{AmSTeX}{}{}%
\def\@@eqncr{\let\@tempa\relax
    \ifcase\@eqcnt \def\@tempa{& & &}\or \def\@tempa{& &}%
      \else \def\@tempa{&}\fi
     \@tempa
     \if@eqnsw
        \iftag@
           \@taggnum
        \else
           \@eqnnum\stepcounter{equation}%
        \fi
     \fi
     \global\tag@false
     \global\@eqnswtrue
     \global\@eqcnt\z@\cr}

\def\TCItag{\@ifnextchar*{\@TCItagstar}{\@TCItag}}
\def\@TCItag#1{%
    \global\tag@true
    \global\def\@taggnum{(#1)}}
\def\@TCItagstar*#1{%
    \global\tag@true
    \global\def\@taggnum{#1}}
\def\dsum{\mathop{\displaystyle \sum }}%
\def\dprod{\mathop{\displaystyle \prod }}%

\RequirePackage{amsmath}
\makeatother
\newcounter{saveeqn}
\newcounter{App} 
\newcommand{\app}{%
\stepcounter{App}%
\setcounter{saveeqn}{\value{equation}}%
\setcounter{equation}{0}%
\renewcommand{\theequation}{\Alph{App}\arabic{equation}} }
\newcommand{\appende}{%
\setcounter{equation}{\value{saveeqn}}%
\renewcommand{\theequation}{\arabic{equation}}  }

\begin{document}

\begin{center}

{\Huge Goldfishing by gauge theory}

\bigskip

\textbf{F. Calogero}$^{a,b,1}$\textbf{\ and E. Langmann}$^{c,2}$

$^{a}$Dipartimento di Fisica, Universit\`{a} di Roma "La Sapienza", I-00185
Roma, Italy

$^{b}$Istituto Nazionale di Fisica Nucleare, Sezione di Roma

$^{c}$Theoretical Physics, KTH, AlbaNova, SE-106 91 Stockholm, Sweden

$^{1}$francesco.calogero@roma1.infn.it, francesco.calogero@uniroma1.it

$^{2}$langmann@kth.se

\bigskip

\bigskip

\textit{Abstract}
\end{center}

A new \textit{solvable} many-body problem of goldfish type is identified and
used to revisit the connection among two different approaches to \textit{%
solvable} dynamical systems. An \textit{isochronous} variant of this model
is identified and investigated. Alternative versions of these models are
presented. The behavior of the alternative \textit{isochronous} model near
its equilibrium configurations is investigated, and a remarkable \textit{%
Diophantine} result, as well as related \textit{Diophantine} conjectures,
are thereby obtained.

\bigskip

\bigskip

\begin{center}
\newpage
\end{center}

\section*{I. Introduction}

Recently a method has been introduced and exploited to identify new \textit{%
exactly solvable} (namely solvable by purely algebraic operations, such as
diagonalizing a matrix) many-body problems characterized by equations of
motion of Newtonian type ("the acceleration of each particle is determined
by the positions and velocities of all particles"), including in particular
models of goldfish type (see, for instance, \cite{FC2001}, and below). The
main idea of this approach -- hereafter referred to as the \textit{direct}
method -- is to start from an \textit{explicitly solvable} matrix evolution
equation (possibly even quite a trivial one), and to then focus on the time
evolution of the \textit{eigenvalues} of this matrix. For an overview of
this method (including an explanation of the terminology used herein), of
the main results yielded by it so far, and the quotation of relevant
references, we refer to the very recent paper \cite{BC2006}.

Another method has been introduced some years ago to treat certain well-know
\ \textit{solvable} dynamical systems and to illuminate their connection
with developments in theoretical particle physics. The main idea of this
approach -- hereafter referred to as the \textit{gauge theory }method -- is
to start from a \textit{gauge invariant} matrix evolution equation and to
exploit the possibility that in one gauge this evolution be trivially simple
hence \textit{solvable} while in another gauge it be related to interesting
evolutions, in particular to the equations of motion of Newtonian type of
certain many-body problems. For an overview of this approach and the
quotation of relevant references, we refer to the relatively recent paper 
\cite{BL1999}. An analysis of the \textit{gauge theory} approach entailing a
clarification of the relation of this method to the \textit{direct} approach
is already provided in the more recent paper published by one of us (EL) 
\cite{L}.

In the present paper, in the context of revisiting this connection, we
identify a new \textit{solvable} many-body problem of goldfish type. This
finding hinges on a result obtained many years ago by V. I. Inozemtsev \cite%
{Ino}. We also present the \textit{isochronous} variant of this many-body
problem, as well as alternative formulations of these two models, and by
investigating the behavior of the alternative \textit{isochronous} model in
the neighborhood of its equilibrium configurations we identify certain
remarkable \textit{Diophantine} relations.

The main \textit{new }results obtained in this paper are reported in the
following Section II. The hasty browser eager to see immediately the
equations of motion of the new \textit{solvable} many-body problems of
goldfish type should jump to (\ref{Gold}) and for the \textit{isochronous}
variant to (\ref{IsoGold}), and for the alternative versions of these models
to (\ref{AltGold}) and (\ref{AltIsoGold}); a \textit{Diophantine} finding
and related conjectures are reported at the end of Section II. In Section
III the \textit{solvable} character of the new many-body problems of
goldfish type is demonstrated, firstly via the \textit{direct} method and
then via the \textit{gauge theory} method; the connection among these two
approaches is thereby illuminated. In Section IV \textit{solvable} dynamical
systems are derived, which constitute nontrivial alternative reformulations
of the many-body problems of goldfish type treated in Section III. In
Section V the behavior of the alternative \textit{isochronous} model in the
neighborhood of its equilibrium configurations is investigated and
remarkable \textit{Diophantine} relations are thereby obtained. In Section
VI possible future developments are mentioned. The Appendix contains some
findings the insertion of which where they are first mentioned (see \textit{%
Remark 2.9} in Section II) would have been too distracting.

\bigskip

\section*{II. Main results}

In this section we report the main \textit{new} findings obtained in
subsequent sections.

The \textit{solvable} $N$-body problem of goldfish type identified in this
paper is characterized by the following equations of motion of Newtonian
type:%
\begin{equation}
\ddot{z}_{n}=2\,z_{n}\,(z_{n}^{\,2}-a^{\,2})+2\,\sum_{m=1,m\neq n}^{N}\frac{%
\left( \dot{z}_{n}+z_{n}^{\,2}-a^{\,2}\right) \,\left( \dot{z}%
_{m}+z_{m}^{\,2}-a^{\,2}\right) }{z_{n}-z_{m}}~.  \label{Gold}
\end{equation}

\textit{Notation}:\textit{\ }$z_{n}\equiv z_{n}(t)$ are the dependent
variables, $t$ is the independent variable ("time"), superimposed dots
denote time-differentiations, $a^{\,2}$ is an \textit{arbitrary} constant
(we use $a^{\,2}$ rather than $a$ merely for notational convenience, see
below), $N$ is a positive integer (generally we assume $N>1),$ and indices
such as $n,m$ generally take all the values $1,2,...,N$ unless otherwise
mentioned.

\textit{Remark 2.1}. Trivially related models involving additional arbitrary
constants could of course be obtained by rescaling the (dependent and
independent) variables and by shifting by a constant amount the dependent
variables; note incidentally that the first factor $2$ in the right-hand
side of (\ref{Gold}) could be changed by rescaling (we put it there for
notational convenience, see below), while the second factor $2$ (that
multiplying the sum) cannot of course be changed. $\boxdot $

\textit{Remark 2.2}. Although for \textit{real} $a^{\,2}$ and for \textit{%
real} initial data $z_{n}(0),$ $\dot{z}_{n}(0)$ the time evolution (for 
\textit{real }time) of this many-body model entails that the dependent
variables $z_{n}(t)$ are as well \textit{real}, we generally assume the time
evolution to take place in the \textit{complex} $z$-plane (and generally
allow the constant $a$ to be as well \textit{complex}); indeed such an
evolution is much more interesting due to the possibility of the "particles"
characterized by the \textit{complex }coordinates $z_{n}(t)$ to go round
each other and the related fact that initial data $z_{n}(0),$ $\dot{z}_{n}(0)
$ leading to particle collisions are then \textit{exceptional} (they
generally have vanishing dimensionality relative to \textit{generic} initial
data). (If attention is instead restricted to \textit{real} motions, then
the trivial change of dependent variables $z_{n}\rightarrow i\,y_{n}$ with $%
y_{n}$ \textit{real} might be expedient in order to deal with \textit{%
confined} motions.) It is possible to reformulate these \textit{complex}
equations of motions as \textit{real }(and even \textit{covariant}, even 
\textit{rotation-invariant}) equations of motion describing the motion of 
\textit{real} point particles in the \textit{real} (say, \textit{horizontal}%
) plane, but we will not take space here to reformulate them in this manner,
since the technique to do so is well-known (see for instance Ref. \cite%
{C2001}). $\boxdot $

The \textit{solvable} character of these equations of motion is evidenced by
the well-known fact \cite{Ino} \cite{C2001} that the $N\times N$ matrix
evolution equation%
\begin{equation}
\ddot{U}=2\,U\,\left( U^{\,2}-a^{\,2}\right)  \label{EqU}
\end{equation}%
is itself \textit{solvable} (in terms of appropriate sigma functions \cite%
{Ino}), together with the following

\textbf{Proposition 2.3}. \textit{The solution of the initial-value problem
for the equations of motion (\ref{Gold}) is provided by the following
prescription: the coordinates }$z_{n}\left( t\right) $\textit{\ are the }$N$%
\textit{\ eigenvalues of the }$N\times N$\textit{\ matrix }$U\left( t\right) 
$\textit{\ solution of (\ref{EqU}) and determined by the following initial
data:} 
\begin{subequations}
\label{Uinitial}
\begin{equation}
U_{nm}(0)=\delta _{nm}\,z_{n}(0)~,  \label{Uinitiala}
\end{equation}%
\begin{eqnarray}
\dot{U}_{nm}(0) &=&-\delta _{nm}\,\left[ z_{n}^{\,2}(0)-a^{\,2}\right]  
\notag \\
&&+\left[ \dot{z}_{n}(0)+z_{n}^{\,2}(0)-a^{\,2}\right] ^{\,1/2}\,\left[ \dot{%
z}_{m}(0)+z_{m}^{\,2}(0)-a^{\,2}\right] ^{\,1/2}~.~\boxdot 
\label{Uinitialb}
\end{eqnarray}%
Note that the matrix $U(0)$\ is \textit{diagonal}, while the matrix $\dot{U}%
(0)$\ is the sum of a \textit{diagonal}\ matrix and a \textit{dyadic}\
matrix.

\textit{Notation}: here and hereafter $\delta _{nm}\equiv \delta _{n,m}$ is
the Kronecker delta symbol, $\delta _{nm}=1$ if $n=m,$ $\delta _{nm}=0$ if $%
n\neq m$.

To obtain the \textit{isochronous} variant of this many-body problem one
starts from the equations of motion 
\end{subequations}
\begin{equation}
\zeta _{n}^{\prime \prime }=2\,\zeta _{n}^{\,3}+2\,\sum_{m=1,m\neq n}^{N}%
\frac{\left( \zeta _{n}^{\prime }+\zeta _{n}^{\,2}\right) \,\left( \zeta
_{m}^{\prime }+\zeta _{m}^{\,2}\right) }{\zeta _{n}-\zeta _{m}}~,
\label{Goldtau}
\end{equation}%
which correspond to (\ref{Gold}) with $a=0$ and with the merely notational
replacement of the dependent variables $z_{n}(t)$ with the dependent
variables $\zeta _{n}\left( \tau \right) $ (and of course now the appended
primes denote differentiations with respect to $\tau $). One can then apply
the procedure usually referred to as "the trick" (see for instance \cite%
{FC2001} \cite{C2001}), i. e. (in this case) the following change of
dependent and independent variables 
\begin{subequations}
\label{Trick}
\begin{equation}
\tilde{z}_{n}(t)=\exp (i\,t)\,\zeta _{n}(\tau )~,  \label{Tricka}
\end{equation}%
\begin{equation}
\tau =i\,\left[ 1-\exp (i\,t)\right] ~.  \label{Trickb}
\end{equation}%
This yields the equations of motion 
\end{subequations}
\begin{equation}
\overset{\cdot \cdot }{\tilde{z}}_{n}=3\,i\,\overset{\cdot }{\tilde{z}}%
_{n}+2\,\tilde{z}_{n}\,(1+\tilde{z}_{n}^{\,2})+2\,\sum_{m=1,m\neq n}^{N}%
\frac{\left( \overset{\cdot }{\tilde{z}}_{n}-i\,\tilde{z}_{n}+\tilde{z}%
_{n}^{\,2}\right) \,\left( \overset{\cdot }{\tilde{z}}_{m}-i\,\tilde{z}_{m}+%
\tilde{z}_{m}^{\,2}\right) }{\tilde{z}_{n}-\tilde{z}_{m}}~.  \label{IsoGold}
\end{equation}

The solution of the initial-value problem is then obviously given by the
solution (via \textbf{Proposition 2.3}) of the problem (\ref{Goldtau}) and
by the "trick" relations (\ref{Trick}), that clearly also imply%
\begin{equation}
\zeta _{n}(0)=\tilde{z}_{n}(0)~,~~~\zeta ^{\prime }(0)=\overset{\cdot }{%
\tilde{z}}_{n}(0)-i\,\tilde{z}_{n}(0)~.  \label{Initial}
\end{equation}%
Equivalently, the solution of this model (\ref{IsoGold}) is clearly given by
the following

\textbf{Proposition 2.4}. \textit{The dependent variables }$\tilde{z}_{n}(t)$%
\textit{\ that solve the initial-value problem for the Newtonian }$N$\textit{%
-body problem (\ref{IsoGold}) are the }$N$\textit{\ eigenvalues of} \textit{%
the }$N\times N$ \textit{matrix }$\tilde{U}(t)$\textit{\ evolving according
to the }solvable\textit{\ matrix evolution equation}%
\begin{equation}
\overset{\cdot \cdot }{\tilde{U}}-3\,i\,\overset{\cdot }{\tilde{U}}-2\,%
\tilde{U}=2\,\tilde{U}^{\,3}  \label{EqUtilde}
\end{equation}%
\textit{and being moreover characterized by the following initial data:} 
\begin{subequations}
\label{UtildeInitial}
\begin{equation}
\tilde{U}_{nm}(0)=\delta _{nm}\,\tilde{z}_{n}(0)~,  \label{UtildeInitiala}
\end{equation}%
\begin{eqnarray}
\overset{\cdot }{\tilde{U}}_{nm}(0)=-\delta _{nm}\,\left[ \tilde{z}%
_{n}^{\,2}(0)\right] &&  \notag \\
+\left[ \overset{\cdot }{\tilde{z}}_{n}(0)-i\,\tilde{z}_{n}(0)+\tilde{z}%
_{n}^{\,2}(0)\right] ^{\,1/2}\,\left[ \overset{\cdot }{\tilde{z}}_{m}(0)-i\,%
\tilde{z}_{m}(0)+\tilde{z}_{m}^{\,2}(0)\right] ^{\,1/2}~.~\boxdot &&
\end{eqnarray}%
Note that the matrix $\tilde{U}(0)$\ is \textit{diagonal}, while the matrix $%
\overset{\cdot }{\tilde{U}}(0)$\ is the sum of a \textit{diagonal}\ matrix
and a \textit{dyadic}\ matrix.

The \textit{solvable} character of the matrix evolution equation (\ref%
{EqUtilde}) is implied by the "trick" formula 
\end{subequations}
\begin{equation}
\tilde{U}(t)=\exp (i\,t)\,U(\tau )~,~~~\tau =i\,\left[ 1-\exp (i\,t)\right] 
\label{TrickU}
\end{equation}%
relating the $N\times N$\ matrix $\tilde{U}(t)$\ evolving according to (\ref%
{EqUtilde}) to the $N\times N$\ matrix $U(t)$\ evolving according to (\ref%
{EqU}) with $a=0$.

\textit{Remark 2.5}. The \textit{solvable} character \cite{Ino} of the
matrix evolution equation (\ref{EqU}) entails that \textit{all} its
solutions $U(t)$ are \textit{meromorphic} functions of the independent
variable $t$. Hence (see (\ref{TrickU})) \textit{all} the \textit{nonsingular%
} solutions $\tilde{U}(t)$ of the matrix evolution equation (\ref{EqUtilde})
are \textit{periodic }with period $2\,\pi ,$%
\begin{equation}
\tilde{U}(t+2\,\pi )=\tilde{U}(t)~.
\end{equation}%
The \textit{singular} solutions of (\ref{EqUtilde}) are \textit{exceptional}%
, corresponding to a set of initial data having \textit{vanishing} measure
with respect to the set of \textit{generic }initial data. $\boxdot $

As an immediate consequence of \textbf{Proposition 2.4} and of this \textit{%
Remark 2.5} there holds the following

\textbf{Proposition 2.6}. All\textit{\ the solutions of the many-body
problem (\ref{IsoGold}) (except those }exceptional\textit{\ ones that run
into a collision of two or more particles, which correspond to }nongeneric%
\textit{\ initial data) are }completely periodic\textit{\ with a period
which is a }positive integer\ multiple\textit{\ }$p$ \textit{of }$2$\textit{%
\thinspace }$\pi $:%
\begin{equation}
\tilde{z}_{n}\left( t+2\,p\,\pi \right) =\tilde{z}_{n}\left( t\right)
~,~~~p=1~\,\text{or \thinspace }2~\,\text{or}...\text{or~~}N~.
\end{equation}%
\textit{The }positive integer $p$ \textit{accounts for the possibility that
the eigenvalues get exchanged among each other through the motion: it
depends on the initial data, but it does not change for sufficiently small,
if finite, changes of these data and it clearly is }not larger\textit{\ than 
}$N$\textit{. }$\boxdot $\textit{\ }

This proposition displays the \textit{isochronous} character of the $N$-body
problem (\ref{IsoGold}), indeed it justifies considering it as one more
instance of \textit{nonlinear harmonic oscillators} \cite{CaIn}.

There exists a, by now rather standard, technique to reformulate these type
of $N$-body problems, by identifying the $N$ "particle coordinates" $%
z_{n}(t) $ as the $N$ zeros of a (\textit{monic}) polynomial in $z$ of
degree $N,$ and by then focusing on the corresponding time evolution of the $%
N$ coefficients $c_{m}(t)$ of this polynomial (see for instance \cite{C1978} 
\cite{C2001}):%
\begin{equation}
\psi (z,t)=\dprod\limits_{n=1}^{N}\left[ z-z_{n}(t)\right]
=\sum_{m=0}^{N}c_{m}(t)\,z^{\,N-m}~,~~~c_{0}=1~.  \label{psi}
\end{equation}
In Section IV we show how such a procedure is applicable in our case, and we
thereby obtain the following alternative formulation of the $N$-body problem
(\ref{Gold}):%
\begin{eqnarray}
&&\ddot{c}_{m}+2\,\left( m-1\right) \,\dot{c}_{m+1}-2\,c_{1}\,\dot{c}%
_{m}+2\,\left( N+1-m\right) \,a^{\,2}\,\dot{c}_{m-1}  \notag \\
&&+\left( m+2\right) \,\left( m-3\right) \,c_{m+2}-2\,\left( m-1\right)
\,c_{1}\,c_{m+1}  \notag \\
&&+2\,\left[ m\,\left( N+2-m\right) \,a^{\,2}+\dot{c}_{1}-c_{1}^{\,2}+3%
\,c_{2}\right] \,c_{m}  \notag \\
&&-2\,\left( N+1-m\right) \,a^{\,2}\,c_{1}\,c_{m-1}+\left( N+2-m\right)
\,\left( N+1-m\right) \,a^{\,4}\,c_{m-2}=0~,  \notag \\
m &=&1,...,N~,~~~c_{0}=1~,~~~c_{-1}=c_{N+1}=c_{N+2}=0~.  \label{AltGold}
\end{eqnarray}

\textit{Remark 2.7}. The ODE of this system with $m=0$ is identically
satisfied; the ODE with $m=N+1$ is also satisfied provided one sets $%
c_{N+3}=0,$ and even the ODE with $m=N+2$ is identically satisfied if one
moreover sets $c_{N+4}=0$. $\boxdot $

\textit{Remark 2.8}. A superficial look at this system of ODEs might suggest
that it is a \textit{linear} system of evolution equations for the
quantities $c_{m}(t);$ but this is of course \textit{not} the case, due to
the presence of the quantities $c_{1}(t)$ and $c_{2}(t)$. Indeed the highly
nonlinear character of this system is already evident by looking at the $N=2$
case, in which case it yields the following (\textit{solvable}!) fourth
order ODE for $f(t)\equiv c_{1}(t)$ :%
\begin{eqnarray}
f^{\prime \prime \prime \prime }\,f^{\,2}-2\,f^{\prime \prime \prime
}\,f^{\prime }\,f^{\,2}-2\,f^{\prime \prime \prime }\,f^{\,3}-2\,\left(
f^{\prime \prime }\right) ^{\,2}\,f+2\,f^{\prime \prime }\,\left( f^{\prime
}\right) ^{\,2}+4\,f^{\prime \prime }\,f^{\prime }\,f^{\,2}-2\,f^{\prime
\prime }\,f^{\,4} &&  \notag \\
-4\,\left( f^{\prime }\right) ^{\,2}\,f^{\,3}+4\,f^{\prime
}\,f^{\,5}+4\,a^{\,2}\,\left( f^{\prime \prime }\,f^{\,2}-2\,f^{\prime
}\,f^{\,3}\right) =0 &&
\end{eqnarray}%
(here for typographical convenience differentiations are denoted by appended
primes rather than superimposed dots). $\boxdot $

\textit{Remark 2.9}. Two equilibrium (namely, time-independent) solutions of
this system (\ref{AltGold}) are provided by the formula%
\begin{equation}
c_{m}=\left( \pm a\right) ^{\,m}\,\binom{N}{m}~.  \label{EquiGoldSpec}
\end{equation}%
They are not, however, the only equilibrium configurations of this system. A
technique to obtain \textit{all} these configurations (including this one!)
is described in the Appendix. $\boxdot $

As indicated above, see (\ref{psi}), the quantities $c_{m}(t)$ that evolve
according to the system of ODEs (\ref{AltGold}) are just the \textit{%
coefficients} of the \textit{monic} polynomial $\psi (z,t)$ of degree $N$ in 
$z,$ the $N$ zeros $z_{n}(t)$ of which evolve according to the equations of
motion (\ref{Gold}). Hence (see \textbf{Proposition 2.3}) the solution of
the system of ODEs (\ref{AltGold}) is given by the following

\textbf{Proposition 2.10}. \textit{The dependent variables }$c_{m}(t)$%
\textit{\ that solve the initial-value problem for the system of nonlinear
ODEs (\ref{AltGold}) are the }$N$\textit{\ coefficients of the polynomial }$%
\psi (z,t),$\textit{\ see (\ref{psi}), which is itself given by the formula}%
\begin{equation}
\psi (z,t)=\det \left[ z-U(t)\right] ~,  \label{psidet}
\end{equation}%
\textit{where the }$N\times N$\textit{\ matrix }$U(t)$\textit{\ evolves
according to the }solvable\textit{\ matrix evolution equation (\ref{EqU})
and is moreover characterized by the initial data (\ref{Uinitial}), with the
initial values }$z_{n}(0)$\textit{, }$\dot{z}_{n}(0)$\textit{\ related to
the initial values }$c_{m}(0),\dot{c}_{m}(0)$\textit{\ by the formulas
implied by (\ref{psi}), } 
\begin{subequations}
\label{Inic}
\begin{equation}
\dprod\limits_{n=1}^{N}\left[ z-z_{n}(0)\right] =\sum_{m=0}^{N}c_{m}(0)\,z^{%
\,N-m}~,~~~c_{0}=1~,  \label{Inica}
\end{equation}%
\begin{equation}
-\sum_{n=1}^{N}\dot{z}_{n}(0)\dprod\limits_{m=1,m\neq n}^{N}\left[ z-z_{n}(0)%
\right] =\sum_{m=1}^{N}\dot{c}_{m}(0)\,z^{\,N-m}~.~\boxdot  \label{Inicb}
\end{equation}

To obtain an alternative version of the \textit{isochronous} $N$-body
problem (\ref{IsoGold}) we use the following version of the "trick": 
\end{subequations}
\begin{subequations}
\label{Trickcm}
\begin{equation}
\tilde{c}_{m}(t)=\left( -i\right) ^{\,m}\,\exp \left( m\,i\,t\right)
\,\gamma _{m}(\tau )~,  \label{Trickcma}
\end{equation}%
\begin{equation}
\tau =i\,\left[ 1-\exp (i\,t)\right] ~.  \label{Trickcmb}
\end{equation}%
Here the quantities $\gamma _{m}\left( \tau \right) $ are the dependent
variables of the previous model, (\ref{AltGold}), with $a=0$, up to the
(purely notational) change consisting in calling the independent variable $%
\tau $ (instead of $t$) and the dependent variables $\gamma _{m}$ (instead
of $c_{m}$), so that these variables satisfy the following system of ODEs: 
\end{subequations}
\begin{eqnarray}
&&\gamma _{m}^{\prime \prime }+2\,\left( m-1\right) \,\gamma _{m+1}^{\prime
}-2\,\gamma _{1}\,\gamma _{m}^{\prime }+\left( m+2\right) \,\left(
m-3\right) \,\gamma _{m+2}  \notag \\
&&-2\,\left( m-1\right) \,\gamma _{1}\,\gamma _{m+1}+2\,\left( \gamma
_{1}^{\prime }-\gamma _{1}^{\,2}+3\,\gamma _{2}\right) \,\gamma _{m}=0 
\notag \\
m &=&1,...,N~,~~~\gamma _{0}=1~,~~~\gamma _{-1}=\gamma _{N+1}=\gamma
_{N+2}=0~,  \label{gammatau}
\end{eqnarray}%
where of course appended primes denote differentiations with respect to the
independent variable $\tau $ (which we allow to be complex, see (\ref%
{Trickcmb})).

Then clearly by applying the "trick" (\ref{Trickcm}) to the system (\ref%
{gammatau}) the following new system of nonlinear ODEs is obtained:%
\begin{eqnarray}
&&\overset{\cdot \cdot }{\tilde{c}}_{m}+2\,\left( m-1\right) \,i\,\overset{%
\cdot }{\tilde{c}}_{m+1}-\left( 2\,m+1+2\,\tilde{c}_{1}\right) \,i\,\overset{%
\cdot }{\tilde{c}}_{m}  \notag \\
&&-\left( m+2\right) \,\left( m-3\right) \,\tilde{c}_{m+2}+2\,\left(
m-1\right) \,\left( m+1+\tilde{c}_{1}\right) \,\tilde{c}_{m+1}  \notag \\
&&+\left[ -m\,\left( m+1\right) +2\,i\,\overset{\cdot }{\tilde{c}}%
_{1}-2\,(m-1)\,\tilde{c}_{1}+2\,\tilde{c}_{1}^{\,2}-6\,\tilde{c}_{2}\right]
\,\tilde{c}_{m}=0~,  \notag \\
&&m=1,...,N~,~~~\tilde{c}_{0}=1~,~~~\tilde{c}_{-1}=\tilde{c}_{N+1}=\tilde{c}%
_{N+2}=0~.  \label{AltIsoGold}
\end{eqnarray}

\textit{Remark 2.11}. The prefactor $\left( -i\right) ^{\,m}$ in (\ref%
{Trickcma}) is of course unessential, it has been introduced merely to give
a marginally nicer look to this system (\ref{AltIsoGold}) and to some other
formulas, see below. With this version, (\ref{Trickcm}), of the "trick" the
relation among the particle coordinates satisfying the equations of motion
of the \textit{isochronous} $N$-body problem (\ref{IsoGold}) and the
quantities $\tilde{c}_{m}(t)$ satisfying this system of ODEs (\ref%
{AltIsoGold}) reads now%
\begin{equation}
\tilde{\psi}(z,t)=\dprod \left[ z-\tilde{z}_{n}(t)\right] =\sum_{m=0}^{N}%
\left( i\right) ^{\,m}\,\tilde{c}_{m}(t)\,z^{\,N-m}~,~~~\tilde{c}_{0}=1~,
\label{Altpsi}
\end{equation}%
see (\ref{Trick}), (\ref{Trickcm}) and (\ref{psi}). Note that we introduced
here the (new) \textit{monic} polynomial $\tilde{\psi}(z,t)$ having as its $%
N $ zeros the $N$ dependent variables $\tilde{z}_{n}(t)$ satisfying (\ref%
{IsoGold}) and as its $N$ coefficients the $N$ dependent variables $\tilde{c}%
_{m}(t)$ satisfying (\ref{AltIsoGold}). $\boxdot $

This model, (\ref{AltIsoGold}), is obviously just as \textit{solvable} as
the previous one, (\ref{AltGold}), indeed the solution of its initial-value
problem can be obtained from the solution of the corresponding problem for (%
\ref{AltGold}) via the formulas (\ref{Trickcm}) that clearly imply the
following relations among the initial data of the two models: 
\begin{subequations}
\begin{equation}
\tilde{c}_{m}(0)=\left( -i\right) ^{\,m}\,\gamma _{m}(0)~,
\end{equation}%
\begin{equation}
\overset{\cdot }{\tilde{c}}_{m}(0)-m\,i\,\tilde{c}_{m}(0)=\left( -i\right)
^{\,m}\,\gamma _{m}(0)~.
\end{equation}%
Equivalently, the solution of this model (\ref{AltIsoGold}) is clearly given
by the following

\textbf{Proposition 2.12}. \textit{The dependent variables }$\tilde{c}_{m}(t)
$\textit{\ that solve the initial-value problem for the system of nonlinear
ODEs (\ref{AltIsoGold}) are the }$N$\textit{\ coefficients of the polynomial 
}$\tilde{\psi}(z,t),$\textit{\ see (\ref{Altpsi}), which is itself given by
the formula} 
\end{subequations}
\begin{equation}
\tilde{\psi}(z,t)=\det \left[ z-\tilde{U}(t)\right] ~,
\end{equation}%
\textit{where the }$N\times N$\textit{\ matrix }$\tilde{U}(t)$\textit{\
evolves according to the }solvable\textit{\ matrix evolution equation} (\ref%
{EqUtilde})\textit{\ and is moreover characterized by the initial data (\ref%
{UtildeInitial}) with the initial values }$\tilde{z}_{n}(0)$\textit{, }$%
\overset{\cdot }{\tilde{z}}_{n}(0)$\textit{\ related to the initial values }$%
\tilde{c}_{m}(0),\overset{\cdot }{\tilde{c}}_{m}(0)$\textit{\ by the
following formulas implied by (\ref{Altpsi}),} 
\begin{subequations}
\label{Inic1}
\begin{equation}
\dprod\limits_{n=1}^{N}\left[ z-\tilde{z}_{n}(0)\right] =\sum_{m=0}^{N}%
\left( i\right) ^{\,m}\,\tilde{c}_{m}(0)\,z^{\,N-m}~,~~~\tilde{c}_{0}=1~,
\end{equation}%
\begin{equation}
-\sum_{n=1}^{N}\left[ \overset{\cdot }{\tilde{z}}_{n}(0)-i\,\tilde{z}_{n}(0)%
\right] \dprod\limits_{m=1,m\neq n}^{N}\left[ z-\tilde{z}_{n}(0)\right]
=\sum_{m=1}^{N}\left( i\right) ^{\,m}\,\overset{\cdot }{\tilde{c}}%
_{m}(0)\,z^{\,N-m}~.~\boxdot 
\end{equation}

As an immediate consequence of this \textbf{Proposition 2.12} and of \textit{%
Remark 2.5 }there holds the following

\textbf{Proposition 2.13}. All \textit{the }nonsingular\textit{\ solutions
of the system of ODEs (\ref{AltIsoGold}) are }completely periodic\textit{\
with period }$2\,\pi $\textit{,} 
\end{subequations}
\begin{equation}
\tilde{c}_{m}(t+2\,\pi )=\tilde{c}_{m}(t)~,  \label{cmper}
\end{equation}%
\textit{while the }singular\textit{\ solutions are }exceptional\textit{,
corresponding to a set of initial data having }vanishing\textit{\ measure
with respect to the set of }generic\textit{\ initial data}. $\boxdot $

This proposition displays the \textit{isochronous} character of the $N$-body
problem (\ref{AltIsoGold}), indeed it justifies considering it as one more
instance of \textit{nonlinear harmonic oscillators} \cite{CaIn}.

Finally, in Section V we obtain \textit{all} the \textit{equilibrium
configurations} of the \textit{isochronous} systems (\ref{IsoGold}) and (\ref%
{AltIsoGold}) and we study the behavior of the system of \textit{nonlinear
harmonic oscillators} (\ref{AltIsoGold}) in the neighborhood of its
equilibrium configurations. The interested reader will find these results in
that section, but we advertise here the \textit{Diophantine} findings
arrived at via this study.

\textbf{Proposition 2.14}. \textit{Let the two }$N\times N$\textit{\
matrices }$A$\textit{\ and }$B$ \textit{be defined componentwise as follows:}
\begin{subequations}
\label{AB}
\begin{equation}
A_{nm}=2\,\left( n-1\right) \,\delta _{n+1,m}-\left( 2\,n+1+2\,\bar{c}%
_{1}\right) \,\delta _{n,m}+2\,\bar{c}_{n}\,\delta _{1,m}~,  \label{ABa}
\end{equation}%
\begin{eqnarray}
B_{nm} &=&\left( n+2\right) \,\left( n-3\right) \,\delta _{n+2,m}-2\,\left(
n-1\right) \,\left( n+1+\bar{c}_{1}\right) \,\delta _{n+1,m}  \notag \\
&&+\left[ n\,\left( n+1\right) +2\,(n-1)\,\bar{c}_{1}-2\,\bar{c}%
_{1}^{\,2}+6\,\bar{c}_{2}\right] \,\delta _{n,m}  \notag \\
&&+2\,\left[ -\left( n-1\right) \,\bar{c}_{n+1}+(n-1-2\,\bar{c}_{1})\,\bar{c}%
_{n}\right] \,\delta _{1,m}+6\,\bar{c}_{n}\,\delta _{2,m}~,  \label{ABb}
\end{eqnarray}%
\textit{with the numbers }$\bar{c}_{m}$\textit{\ defined as follows:} 
\end{subequations}
\begin{subequations}
\label{cmbar}
\begin{equation}
\text{for }\nu =0~,~~~\bar{c}_{m}=\left( -\right) ^{\,m}\,\binom{\mu }{m}~,
\label{cmbara}
\end{equation}%
\begin{eqnarray}
\text{for }\nu  &=&1~,~~~\bar{c}_{m}=\delta _{0m}+\delta _{1m}~~~\text{if }%
\mu =1~,  \notag \\
\bar{c}_{m} &=&\left( -\right) ^{\,m}\,\left[ \binom{\mu -2}{m}-\binom{\mu -2%
}{m-2}\right] ~~~\text{if }\mu >1~,  \label{cmbarb}
\end{eqnarray}%
\begin{equation}
\text{for }\nu =3~,~~\,\bar{c}_{m}=\left( -\right) ^{\,m}\,\left[ \binom{\mu
-3}{m}+6\,\binom{\mu -3}{m-1}+14\,\binom{\mu -3}{m-2}+14\,\binom{\mu -3}{m-3}%
\right] ~,  \label{cmbarc}
\end{equation}%
\begin{equation}
\text{for }\nu =4\text{ },~~\,\bar{c}_{m}=\left( -\right)
^{\,m}\,\sum_{k=0}^{4}\binom{\mu -4}{m-k}\,\binom{5}{k}~,  \label{cmbard}
\end{equation}%
\begin{equation}
\text{for }\nu =5\text{ },~~\,\bar{c}_{m}=\left( -\right) ^{\,m}\,\left[ c\,%
\binom{\mu -5}{m-5}+\sum_{k=0}^{5}\binom{\mu -5}{m-k}\,\binom{5}{k}\right]
~,~~\,c\text{ \thinspace \thinspace arbitrary~}.  \label{cmbarf}
\end{equation}%
\textit{As indicated above the parameter }$\nu $\textit{\ (the role of which
here is mainly to distinguish }$5$ \textit{different cases) can take any one
of the }$5$\textit{\ values }$0,1,3,4,5$\textit{, while the parameter }$\mu $%
\textit{\ can take any }positive integer\textit{\ value in the range }$\nu
\leq \mu \leq N.$\textit{\ Let the }$2\,N$ \textit{numbers }$p_{n}^{\left(
\pm \right) }$\textit{\ be the eigenvalues of the generalized eigenvalue
problem} 
\end{subequations}
\begin{subequations}
\begin{equation}
\left( p^{\,2}+A\,p+B\right) \,\underline{r}=0~,  \label{Eigena}
\end{equation}%
\textit{(where }$\underline{r}\equiv \left( r_{1},...,r_{N}\right) $\textit{%
\ denotes the corresponding eigenvector) implying}%
\begin{equation}
\det \left( p^{\,2}+A\,p+B\right) =\dprod\limits_{n=1}^{N}\left[ \left(
p-p_{n}^{\left( +\right) }\right) \,\left( p-p_{n}^{\left( -\right) }\right) %
\right] ~.  \label{Eigenb}
\end{equation}%
\textit{Then the }$2\,N$\textit{\ numbers }$p_{n}^{\left( \pm \right) }$%
\textit{\ are }all integers\textit{.} $\boxdot $

\textit{Notation}: here and throughout the symbol $\binom{x}{y}$ is the
standard binomial coefficient, 
\end{subequations}
\begin{equation}
\binom{x}{y}=\frac{\Gamma (x+1)}{\Gamma (y+1)\,\Gamma (x-y+1)}~.
\end{equation}

We have verified with the help of symbolic programing languages (we
used Maple and Mathematica) and for an ample sample of values of $N$
and of the other parameters the validity of this proposition (proven
in Section V), and from these computer-aided checks we are led to
formulate the following \textit{Diophantine}\textbf{\ }conjectures.

\textbf{Conjecture 2.15}. \textit{For }$\nu =0,1,3,4,5$\textit{\ and }$\mu $%
\textit{\ integer in the range }$\nu \leq \mu \leq N$\textit{\ the
eigenvalues of the generalized eigenvalue problem (\ref{Eigena}) (with (\ref%
{AB}) and (\ref{cmbar})) are given by the following formulas:} 
\begin{subequations}
\label{Conj}
\begin{eqnarray}
\text{for }\nu =0~,~~~\det \left( p^{\,2}+A\,p+B\right) &&  \notag \\
=\left\{ \dprod\limits_{n=1}^{N-\mu }\left[ \left( p-n\right) \,\left(
p-n-1\right) \right] \right\} \,\left\{ \dprod\limits_{n=1}^{\mu }\left[
\left( p+n\right) \,\left( p+n-5\right) \right] \right\} ~, &&  \label{Conja}
\end{eqnarray}%
\begin{eqnarray}
\text{for }\nu =1~,~~~\det \left( p^{\,2}+A\,p+B\right) =\left( p+1\right)
\,\left( p-4\right) \cdot &&  \notag \\
\cdot \left\{ \dprod\limits_{n=1}^{N-\mu }\left[ \left( p-n\right) \,\left(
p-n-5\right) \right] \right\} \,\left\{ \dprod\limits_{n=1}^{\mu -1}\left[
\left( p+n\right) \,\left( p+n-7\right) \right] \right\} ~, &&  \label{Conjb}
\end{eqnarray}%
\begin{eqnarray}
\text{for }\nu =3~,~~~\det \left( p^{\,2}+A\,p+B\right) =\left( p+1\right)
\,\left( p-4\right) \cdot &&  \notag \\
\cdot \left\{ \dprod\limits_{n=1}^{N-\mu }\left[ \left( p-n\right) \,\left(
p-n+5\right) \right] \right\} \,\left\{ \dprod\limits_{n=1}^{\mu -1}\left[
\left( p+n\right) \,\left( p-n+\mu -7\right) \right] \right\} ~, &&
\label{Conjc}
\end{eqnarray}%
\begin{eqnarray}
\text{for }\nu =4~,~~~\det \left( p^{\,2}+A\,p+B\right) =\left( p+1\right) \,%
\left[ \dprod\limits_{n=1}^{3}\left( p-n-1\right) \right] \cdot &&  \notag \\
\cdot \left\{ \dprod\limits_{n=1}^{N-\mu }\left[ \left( p-n\right) \,\left(
p-n+1\right) \right] \right\} \,\left[ \dprod\limits_{n=1}^{\mu -4}\left(
p+n\right) \right] \,\left[ \dprod\limits_{n=1}^{\mu }\left( p+n+1\right) %
\right] ~, &&  \label{Conjd}
\end{eqnarray}%
\begin{eqnarray}
\text{for }\nu =5~,~~~\det \left( p^{\,2}+A\,p+B\right) = &&  \notag \\
\cdot \left\{ \dprod\limits_{n=1}^{N-\mu }\left[ \left( p-n\right) \,\left(
p-n-1\right) \right] \right\} \,\left\{ \dprod\limits_{n=1}^{\mu }\left[
\left( p+n\right) \,\left( p-n+\mu -4\right) \right] \right\} ~. &&
\label{Conje}
\end{eqnarray}%
\textit{Here we use the standard convention according to which a product
equals }unity\textit{\ if the lower limit of the running index exceeds the
upper limit. }$\boxdot $

\textit{Remark 2.16.} For $\mu =\nu =0$ the validity of this conjecture is
certainly true, indeed trivially so (see below the \textit{Remark 5.4}). $%
\boxdot $

The \textbf{Conjecture 2.15} only refers to \textit{integer} values of the
parameter $\mu $ in the range $\nu \leq \mu \leq N$. But our computer-aided
exploration also indicates the validity, for \textit{arbitrary} values of
the parameter $\mu $, of the following conjecture (which is only formulated
below for sufficiently large values of $N,$ to avoid less interesting
complications).

\textbf{Conjecture 2.17}. \textit{The} \textit{generalized eigenvalue
problem (\ref{Eigena})\ (with (\ref{AB}) and (\ref{cmbara})) features, for }%
arbitrary $\mu ,$ \textit{the }$N-1$ \textit{eigenvalues} 
\end{subequations}
\begin{subequations}
\label{Conj2}
\begin{equation}
2,\,3,\,4,\,5-\mu ,\,6-\mu ,...,\,N-\mu ~,~~\,\text{\textit{if} }\nu =0~%
\text{\textit{or }}\nu =5~\,\text{\textit{and} }N\geq 5~,  \label{Conj2a}
\end{equation}%
\begin{equation}
2,\,3,\,4,\,4-\mu ,5-\mu ,...,N-1-\mu ~,~~\,\text{\textit{if} }\nu =4~\,%
\text{\textit{and} }N\geq 5~,
\end{equation}%
\textit{and the }$N-4$ \textit{eigenvalues}%
\begin{equation}
-1,\,4,\,6,\,8-\mu ,9-\mu ,...,N-\mu ~,~~\,\text{\textit{if} }\nu =1~\,\text{%
\textit{and} }N\geq 8~,
\end{equation}%
\begin{equation}
-1,\,4,\,6,\,3-\mu ,4-\mu ,...,N-5-\mu ~,~~\,\text{\textit{if} }\nu =3~\,%
\text{\textit{and} }N\geq 8~.~\boxdot
\end{equation}

\textit{Remark 2.18}. The \textbf{Conjecture 2.17} -- in contrast to the 
\textbf{Conjecture 2.15 -- }does \textit{not }provide the \textit{entire}
spectrum of the eigenvalue problem (\ref{Eigena}), which of course features $%
2\,N$ eigenvalues. $\boxdot $

Some aspects of these conjectures are easy to prove. For instance \textbf{%
Conjecture 2.15} can be proven by induction for all $N>\mu $
if one assumes its validity for $N=\mu .$  But complete
proofs of them do not seem quite trivial (see Section VI).

\bigskip

\section*{III. Two proofs of Proposition 2.3}

In this section (part of) the results reported in the preceding Section II
are proven, firstly by the \textit{direct} method, then by the \textit{gauge
theory} method.

\bigskip

\subsection*{A. Direct method}

The starting point of the \textit{direct} method is the \textit{solvable} $%
N\times N$ matrix evolution equation (\ref{EqU}). We then introduce the
eigenvalues of the matrix $U\left( t\right) $ and the corresponding
diagonalizing matrix $R(t)$ via the formulas 
\end{subequations}
\begin{subequations}
\label{URZ}
\begin{equation}
U\left( t\right) =R(t)\,Z(t)\,\left[ R(t)\right] ^{-1}~,  \label{URZa}
\end{equation}%
\begin{equation}
Z\left( t\right) =\text{diag}\left[ z_{n}(t)\right] ~,  \label{URZb}
\end{equation}%
with moreover%
\begin{equation}
R(0)=\mathbf{1}~.  \label{URZc}
\end{equation}%
Here and below $\mathbf{1}$ is the $N\times N$ identity matrix. Note that
the first two of these equations, (\ref{URZa}) and (\ref{URZb}), identify
(consistently with \textbf{Proposition 2.3}) the $N$ coordinates $z_{n}(t)$
as the $N$ eigenvalues of the $N\times N$ matrix~$U(t),$ while the third, (%
\ref{URZc}), is consistent via the first two with the assignment (\ref%
{Uinitiala}).

It is then easy to see (for the derivation of these formulas see, if need
be, for instance \cite{BC2006}) that, after introducing the $N\times N$
matrix $M\left( t\right) $ via the assignment 
\end{subequations}
\begin{equation}
M(t)=\left[ R\left( t\right) \right] ^{-1}\,\dot{R}\left( t\right) ~,
\label{M}
\end{equation}%
one gets 
\begin{subequations}
\begin{equation}
\dot{U}(t)=R(t)\,\left\{ \dot{Z}(t)+\left[ M(t),\,Z(t)\right] \right\} \,%
\left[ R\left( t\right) \right] ^{-1}~,  \label{Udotinitial}
\end{equation}%
entailing (see (\ref{URZc}))%
\begin{equation}
\dot{U}(0)=\dot{Z}(0)+\left[ M(0),\,Z(0)\right] ~,  \label{Udotinitialb}
\end{equation}%
as well as the following system of evolution ODEs for the coordinates $%
z_{n}(t)$ and for the matrix elements $M_{nm}(t)$ of the matrix $M(t)$: 
\end{subequations}
\begin{subequations}
\label{Ev}
\begin{equation}
\ddot{z}_{n}=2\,z_{n}\,(z_{n}^{\,2}-a^{\,2})+2\,\sum_{m=1,m\neq n}^{N}\left(
z_{n}-z_{m}\right) \,M_{nm}\,M_{mn}~,  \label{Eva}
\end{equation}%
\begin{eqnarray}
\frac{\dot{M}_{nm}}{M_{nm}}=-2\,\frac{\dot{z}_{n}-\dot{z}_{m}}{z_{n}-z_{m}}%
-M_{nn}+M_{mm} &&  \notag \\
+\sum_{\ell =1;\ell \neq n,m}^{N}\frac{z_{n}+z_{m}-2\,z_{\ell }}{z_{n}-z_{m}}%
\,\frac{M_{n\ell }\,M_{\ell m}}{M_{nm}}~,~~~n\neq m~. &&  \label{Evb}
\end{eqnarray}%
Note that the time evolutions of the \textit{diagonal} elements $M_{nn}(t)$
of the matrix $M(t)$ remain unrestricted: it is indeed clear from (\ref{URZ}%
) (implying that $R(t)$ is defined only up to multiplication from the right
by an \textit{arbitrary} diagonal matrix $D(t)$) and from (\ref{M}) that
these $N$ functions of time can be chosen arbitrarily without affecting the
eigenvalues of $U(t),$ namely the coordinates $z_{n}(t).$ Indeed it is clear
that the $N\times N$ matrix evolution equation (\ref{EqU}), characterizing
the time evolution of the $N^{\,2}$ matrix elements $M_{nm}(t)$, has now
been turned into the system (\ref{Ev}), characterizing the time evolution of
the $N$ coordinates $z_{n}(t)$ and the $N\,(N-1)$\textit{\ off-diagonal}
elements $M_{nm}(t)$ (with $n\neq m$) of the $N\times N$ matrix $M(t).$

Clearly this system (\ref{Ev}) is \textit{no less solvable} than the
original matrix evolution (\ref{EqU}), because its solution can be retrieved
from the solution of (\ref{EqU}) by purely algebraic operations
(essentially, by diagonalizing an $N\times N$ matrix).

But we are interested in obtaining an $N$-body problem involving \textit{only%
} the $N$ "particle coordinates" $z_{n}(t)$, hence our next task is to
eliminate the $N\,\left( N-1\right) $ "auxiliary quantities" $M_{nm}(t)$
(with $n\neq m)$. To do this one must find (assuming it exists) an
appropriate \textit{ansatz }expressing the $N\,\left( N-1\right) $ auxiliary
quantities $M_{nm}(t)$ (with $n\neq m$) in terms of the $N$ particle
coordinates $z_{n}(t),$ taking advantage if need be of the freedom to assign
the $N$ quantities $M_{nn}(t)$ at our convenience.

An \textit{ansatz }that works (in the sense of turning the $N\,\left(
N-1\right) $ evolution equations (\ref{Evb}) into identities) is 
\end{subequations}
\begin{subequations}
\label{Ans}
\begin{equation}
M_{nn}(t)=-\sum_{\ell =1}^{N}\frac{g}{\left[ z_{n}\left( t\right) -z_{\ell
}\left( t\right) \right] ^{\,2}}~,  \label{Ansa}
\end{equation}%
\begin{equation}
M_{nm}(t)=\frac{g}{\left[ z_{n}\left( t\right) -z_{m}\left( t\right) \right]
^{\,2}}~,~~~n\neq m~,  \label{Ansb}
\end{equation}%
with $g$ an \textit{arbitrary constant}. This leads however to an $N$-body
model the \textit{solvable} character of which is already well known \cite%
{InoIno} \cite{SW} \cite{BI} \cite{V1987}, hence we do not pursue this
development here (we elaborate this point a little further in the following
version of the proof).

\textit{Remark 3.1}$.$ Clearly insertion of this \textit{ansatz} (\ref{Ans})
in the more general matrix evolution equation 
\end{subequations}
\begin{subequations}
\label{GenMat}
\begin{equation}
\ddot{U}=\Phi \left( U\right) ~,  \label{GenMata}
\end{equation}%
with $\Phi (z)$ an \textit{arbitrary} scalar function would also work (since
this does not depend on the equations of motion satisfied by the coordinates 
$z_{n}$, see below), and it would lead to the $N$-body problem characterized
by the Newtonian equations of motion%
\begin{equation}
\ddot{z}_{n}=\Phi \left( z_{n}\right) -2\,\sum_{n=1,m\neq n}^{N}\frac{g^{\,2}%
}{\left( z_{n}-z_{m}\right) ^{\,3}}~.
\end{equation}%
This was already noted, many years ago, by Veselov \cite{V1987}. But it
appears that, so far, the most general (up to trivial transformations) 
\textit{solvable }$N\times N$ matrix evolution of type (\ref{GenMata}) is
just (\ref{EqU}). $\boxdot $

Another \textit{ansatz} that also does (as it were \textit{miraculously})
work (namely, transform the evolution equations (\ref{Evb}) into identities)
reads as follows: 
\end{subequations}
\begin{equation}
M_{nm}(t)=-\frac{\left\{ \left[ \dot{z}_{n}+z_{n}^{\,2}-a^{\,2}\right] \,%
\left[ \dot{z}_{m}+z_{m}^{\,2}-a^{\,2}\right] \right\} ^{\,1/2}}{z_{n}-z_{m}}%
~,~\ \ n\neq m~.  \label{Ansatz}
\end{equation}%
Note that this \textit{ansatz,} in contrast to the previous one, contains no
arbitrary ("coupling") constant $g$. In this case the appropriate assignment
for the diagonal elements $M_{nn}(t)$ is quite trivial: $M_{nn}(t)=0,$ or
equivalently (see (\ref{Evb})) $M_{nn}(t)=\mu (t),$ $\mu (t)$ being an
arbitrary function of time (but independent of the index $n$). The truth of
this assertion can be verified by a trivial if tedious calculation: note
that the evolution equations (\ref{Eva}) must also be used in the process.

And it is now clear that the insertion of this \textit{ansatz} in (\ref{Eva}%
) yields (\ref{Gold}), while its insertion in (\ref{Udotinitialb}) yields
the assignment (\ref{Uinitialb}). The proof of \textbf{Proposition 2.3} is
thus completed.

\bigskip

\subsection*{B. Gauge theory approach}

Let us now prove again \textbf{Proposition 2.3}, but via the \textit{gauge
theory} method. Although this entails some repetitions we believe it is
useful to go through this exercise in some detail, especially because we
will now use a somewhat different language -- and one purpose of this paper
is precisely to clarify the relations among these two different approaches.
Moreover this presentation provides some indication of the extent to which
this kind of fishing expeditions are likely to yield new goldfishes, namely
new interesting \textit{solvable} models.

Let us start by reviewing (but in a notation more conducive to a direct
comparison with the preceding treatment) the gauge theory approach to a more
standard model referred to in the literature as rCM \cite{BL1999} \cite{L},
characterized by the Newtonian equations of motion%
\begin{equation}
\ddot{z}_{n}=-z_{n}-2\,\sum_{m=1,m\neq n}^{N}\frac{g^{\,2}}{\left(
z_{n}-z_{m}^{{}}\right) ^{\,3}}~.  \label{rCM}
\end{equation}%
We then present a variant of this approach which leads us to the new \textit{%
solvable }$N$-body model of goldfish type presented above.

We start with the following system of dynamical equations for the $N\times N$
matrices $U\equiv U(t)$, $M\equiv M(t)$, and $W\equiv W(t)$: 
\begin{subequations}
\label{gauge}
\begin{equation}
\dot{U}+[M,U]=W~,  \label{gauge1}
\end{equation}%
\begin{equation}
\dot{W}+[M,W]=\Phi (U)~,  \label{gauge2}
\end{equation}%
with the square brackets indicating matrix commutators. As discussed below,
to get the rCM model one should assign the function $\Phi (U)$ as follows: 
\end{subequations}
\begin{equation}
\Phi (U)=-U~.  \label{p}
\end{equation}%
It is however convenient to leave this function $\Phi $ unspecified for the
moment; but we require that it contain no other matrix besides $U,$ so that $%
R^{-1}\,\Phi (U)\,R=\Phi (R^{-1}\,U\,R)$ for any (invertible) matrix $R$.

The equations (\ref{gauge}) have a natural interpretation as gauge theory in 
$0+1$ dimensions: they are indeed of the form $[D_{t},U]=P$, $[D_{t},P]=\Phi
(U),$ with $D_{t}=\partial _{t}+M$ being the so-called \textit{covariant
time derivative }with $M$ regarded as \textit{gauge field}. In particular,
they are invariant under the following \textit{gauge transformations,} 
\begin{subequations}
\begin{eqnarray}
U &\rightarrow &\tilde{U}=R^{-1}\,U\,R~,~~~W\rightarrow \tilde{W}%
=R^{-1}\,W\,R~,  \label{GTa} \\
M &\rightarrow &\tilde{M}=R^{-1}\,M\,R+R^{-1}\,\dot{R}~,  \label{GTb}
\end{eqnarray}%
where the matrix $R\equiv R(t)$ characterizing the gauge transformation is
an arbitrarily time-dependent invertible matrix (the transformation rule for 
$M$ follows from $R^{-1}\,D_{t}\,R=\partial _{t}+R^{-1}\,\dot{R}+R^{-1}\,M\,R
$). One can exploit this invariance to impose additional conditions. In
particular for any solution $U(t)$, $M(t)$, and $W(t)$ of (\ref{gauge}), one
can find a gauge transformation $R(t)$ such that the gauge-transformed
matrix $\tilde{M}(t)$, see (\ref{GTb}), vanishes, $\tilde{M}(t)=0$. Indeed,
this is implied by the fact that the linear first-order matrix ODE 
\end{subequations}
\begin{subequations}
\begin{equation}
\dot{R}+M\,R=0,\quad R(0)=\mathbf{1}
\end{equation}%
always has a (unique) solution, which can be written as 
\begin{equation}
R(t)=\mathcal{T}\exp \left[ -\int_{0}^{t}ds\,M(s)\right] 
\end{equation}%
where the symbol $\mathcal{T}$ denotes time-ordering. This shows that we can
impose the condition 
\end{subequations}
\begin{equation}
M(t)=0  \label{Weyl}
\end{equation}%
for all times $t$, without loss of generality. Note that this also implies
that, if we impose (\ref{Weyl}), we will not loose any solution: indeed any
solution to our gauge theory equations can be obtained by solving these
equations with the condition (\ref{Weyl}) imposed, and performing a gauge
transformation afterwards. In particle physics this latter condition is
often called \textit{Weyl gauge}. We will also use another gauge condition,
namely 
\begin{equation}
U_{nm}(t)=\delta _{nm}\,z_{n}(t)  \label{DCG}
\end{equation}%
for all times $t$, which we call \textit{diagonal Coulomb gauge}. Note that
we can impose this latter condition if the matrix $U(t)$ is such that there
exists an invertible matrix $R(t)$ such that $R^{-1}(t)\,U(t)\,R(t)$ is a 
\textit{diagonal} matrix, and this is obviously true in the generic case
when the matrix $U(t)$ is non-degenerate. The cases when the matrix $U(t)$
is degenerate correspond to particles in our $N$-body system colliding, and
then our solution breaks down (as it should): but this can only happen for
exceptional -- i. e., \textit{nongeneric} -- initial data, if we allow the
particle coordinates $z_{n}(t)$ to move in the complex plane, as we
generally do (see the \textit{Remark 2.2}).

The idea now is that, by imposing the gauge condition (\ref{Weyl}), we get
the matrix equation $\ddot{U}=\Phi (U)$ which might be (chosen to be)
exactly solvable, whereas by imposing the condition (\ref{DCG}), we get a
(hopefully) interesting dynamical system for the variables $z_{n}(t)$. The
latter dynamical system can then be solved explicitly as follows: one first
determines the solution $U(t)$ of the matrix equation obtained from (\ref%
{gauge}) in the Weyl gauge and with the initial conditions 
\begin{equation}
U_{nm}(0)=\delta _{nm}\,z_{n}(0),\quad \dot{U}_{nn}(0)=\dot{z}_{n}(0).
\label{IC}
\end{equation}%
Then the eigenvalues of $U(t)$ give the solution of the dynamical system.
Note that we can only assign the \textit{diagonal} elements of $\dot{U}(0)$
since, as we will see, the \textit{off-diagonal} elements of $\dot{U}(0)$
are determined by another conditions which we have to add.

Indeed, to get an interesting dynamical system, we need to add one more
gauge invariant equation to (\ref{gauge}). In particular, to get the rCM
model (\ref{rCM}), one has to add the equation 
\begin{equation}
\lbrack W,U]=J  \label{GaussLaw}
\end{equation}%
which is often called \textit{Gauss law} or \textit{momentum map}. This
latter equation is gauge invariant if the matrix $J=J(t)$ introduced here
transforms under gauge transformations as $J\rightarrow \tilde{J}%
=R^{-1}\,J\,R$. It turns indeed out \cite{BL1999} \cite{L} that if one makes
the assignment%
\begin{equation}
J_{nm}(0)=g\,(1-\delta _{nm})  \label{J0}
\end{equation}%
and chooses $\Phi (U)$ as in (\ref{p}), then the coordinates $z_{n}(t)$ obey
the equations of motion of the rCM model (\ref{rCM}).

To obtain goldfish type dynamical systems one must instead replace the
Gauss' law condition (\ref{GaussLaw}) by 
\begin{subequations}
\label{BB}
\begin{equation}
B_{nm}\,B_{n^{\prime }m^{\prime }}=B_{nm^{\prime }}\,B_{n^{\prime }m}
\label{BBa}
\end{equation}%
with%
\begin{equation}
B=W+f(U)~,  \label{BBb}
\end{equation}%
where $f(x)$ is another function to be assigned later. To see that this
condition is gauge invariant we note that it can be written as $B\otimes B=%
\mathcal{P\,}B\otimes B\,\ $where $\otimes $ is the tensor product (so that $%
(B\otimes B)\,(\,u\otimes v)=\left( B\,u\right) \otimes \left( B\,v\right) $
where $u$ and $v$ are $N$-vectors) and $\mathcal{P}$ is the permutation
matrix defined as follows, $\mathcal{P\,}u\otimes v=v\otimes u$: the gauge
invariance of (\ref{BBa}) follows from the obvious fact that $\mathcal{P}$
commutes with $R\otimes R$.

A new finding (proven below) is then given by the following

\textbf{Proposition 3.2}. \textit{The gauge theory equations (\ref{gauge})
and (\ref{BB}) are consistent if } 
\end{subequations}
\begin{subequations}
\label{f_Phi}
\begin{equation}
f(x)=\alpha +\beta \,x+\gamma \,x^{\,2}~,  \label{f_Phia}
\end{equation}%
\begin{equation}
\Phi (x)=(\alpha +\beta \,x+\gamma \,x^{\,2})\,(\beta +2\,\gamma
\,x)=f(x)\,f^{\prime }(x)~,  \label{f_Phib}
\end{equation}%
\textit{for arbitrary constants }$\alpha ,\beta ,\gamma $\textit{. Imposing
the diagonal Coulomb gauge (\ref{DCG}) these equations imply } 
\end{subequations}
\begin{equation}
\ddot{z}_{n}=\Phi \left( z_{n}\right) +\sum_{m=1,m\neq n}^{N}\frac{[\dot{z}%
_{n}+f(z_{n})]\,[\dot{z}_{m}+f(z_{m})]}{z_{n}-z_{m}}~,  \label{gold}
\end{equation}%
\textit{and thus the solution of the initial-value problem for this
dynamical system can be obtained by solving the corresponding gauge theory
equations in the Weyl gauge, see (\ref{Weyl}). More specifically: the
solution of the initial-value problem for this dynamical system, (\ref{gold}%
), is given by the eigenvalues of the matrix equation} 
\begin{equation}
\ddot{U}=\Phi (U)
\end{equation}%
\textit{with the initial conditions }%
\begin{equation}
U_{nm}(0)=\delta _{mn}\,z_{n}(0)
\end{equation}%
\textit{and} 
\begin{subequations}
\label{Udotinit}
\begin{equation}
\dot{U}_{nn}(0)=\dot{z}_{n}~,  \label{Udotinita}
\end{equation}%
\begin{equation}
\dot{U}_{nm}(0)=\left\{ [\dot{z}_{n}(0)+f\left[ z_{n}(0)\right] \right\}
^{\,1/2}\,\left\{ [\dot{z}_{m}(0)+f\left[ z_{m}(0)\right] \right\}
^{\,1/2}~,~~~n\neq m~.~\boxdot  \label{Udotinitb}
\end{equation}

\textit{Remark 3.3. }In the special case $\gamma =0$ a model is obtained
whose solvability was already known \cite{CI}. Our new model reported in
Section II (and already derived by our other method in the preceding 
Subsection A) is obtained for $\beta =0$, $\gamma =1$, and $\alpha =-a^{2}$. But
the greater generality of the result as formulated in \textbf{Proposition 3.2%
} is only apparent: if $\gamma \neq 0$, one can always reduce this more
general case to the special case with $\beta =0$ by the (rather trivial)
transformations $U\rightarrow \check{U}=U-\frac{\beta }{2\,\gamma }\mathbf{1,%
}$ $t\rightarrow \check{t}=\gamma \,t.$ And note that, for $%
f(z)=z^{\,2}-a^{\,2}$, the right-hand sides of (\ref{Udotinit}) and (\ref%
{Uinitialb}) coincide: the apparent differences are merely notational. $%
\boxdot $

Let us end this section by outlining the proof of this result, whose analogy
with that proven in the first part of this section is we trust evident
enough not to require further elaboration. We firstly write out (\ref{gauge}%
) and (\ref{BB}) by imposing the condition (\ref{Weyl}). Then (\ref{gauge1})
becomes 
\end{subequations}
\begin{equation}
\delta _{nm}\,\dot{z}_{n}+M_{nm}\,(z_{m}-z_{n})=W_{nm}
\end{equation}%
which for the \textit{diagonal} elements (i.e.\ $n=m$) implies 
\begin{equation}
W_{nn}=\dot{z}_{n}~.  \label{qdot}
\end{equation}%
For the \textit{off-diagonal} elements we obtain the assignment%
\begin{equation}
M_{nm}=-\frac{1}{z_{n}-z_{m}}\,W_{nm}~,~~~n\neq m~,  \label{MM}
\end{equation}%
whereas the diagonal elements $M_{nn}$ remain unassigned. Then (\ref{gauge2}%
) reads%
\begin{equation}
\dot{W}_{nm}+\sum_{\ell =1}^{N}\left( M_{n\ell }\,W_{\ell m}-W_{n\ell
}\,M_{\ell m}\right) =\delta _{nm}\,\Phi (z_{n})~.  \label{Pdot}
\end{equation}%
The \textit{diagonal} elements of this equation give (via (\ref{MM}) and (%
\ref{qdot})) 
\begin{equation}
\ddot{z}_{n}-2\sum_{\ell =1,\ell \neq n}^{N}\frac{W_{n\ell }\,W_{\ell n}}{%
z_{n}-z_{\ell }}=\Phi (z_{n}),  \label{pdot}
\end{equation}%
while the \textit{off-diagonal} elements imply the following important
consistency conditions, 
\begin{equation}
\begin{split}
W_{nm}& +(M_{n}-M_{m})\,\dot{W}_{nm}+\frac{\dot{z}_{n}-\dot{z}_{m}}{%
z_{n}-z_{m}}\,W_{nm} \\
& -\sum_{\ell =1,\ell \neq n,m}^{N}W_{n\ell }\,W_{\ell m}\,\left( \frac{1}{%
z_{n}-z_{\ell }}-\frac{1}{z_{\ell }-z_{m}}\right) =0~,\quad n\neq m.
\end{split}
\label{conc}
\end{equation}%
In particular (\ref{BBa}) implies $B_{nm}\,B_{mn}=B_{nn}\,B_{mm}$. Inserting
in this equation the assignment $B_{nm}=W_{nm}+\delta _{nm}\,f(z_{n})$ (see (%
\ref{BBb})) and in particular $B_{nn}=\dot{z}_{n}+f(z_{n})$ (see (\ref{qdot}%
)), we get 
\begin{equation}
W_{nm}\,W_{mn}+\delta _{nm}\,\left[ 2\,\dot{z}_{n}+f(z_{n})\right]
\,f(z_{n})=\left[ \dot{z}_{n}+f(z_{n})\right] \,\left[ \dot{z}_{m}+f(z_{m})%
\right] ~.
\end{equation}%
Thus the solution of (\ref{BB}) for $n^{\prime }=m$ and $m^{\prime }=n$ is 
\begin{equation}
W_{nm}=\exp (\varphi _{n}-\varphi _{m})\,\left\{ -\delta _{nm}\,\left[ 2\,%
\dot{z}_{n}+f(z_{n})\right] \,f(z_{n})+\left[ \dot{z}_{n}+f(z_{n})\right] \,%
\left[ \dot{z}_{m}+f(z_{m})\right] \right\} ^{\,1/2}  \label{BB1}
\end{equation}%
where the functions $\varphi _{n}(t)$ are \textit{arbitrary}, and it is easy
to check that this is a solution of (\ref{BB}) also for all the other values
of $n^{\prime },m^{\prime }$. Inserting this in (\ref{pdot}) we obtain 
\begin{equation}
\ddot{z}_{n}=\Phi (z_{n})+2\sum_{\ell =1,\ell \neq n}^{N}\frac{\left[ \dot{z}%
_{n}+f(z_{n})\right] \,\left[ \dot{z}_{\ell }+f(z_{\ell })\right] }{%
z_{n}-z_{\ell }}~.  \label{ddotzn}
\end{equation}%
\qquad 

It remains to check the consistency relations (\ref{conc}). We note that,
for $\ell \neq n,m$ and $n\neq m$, (\ref{BB1}) entails $W_{n\ell }\,W_{\ell
m}=W_{nm}\,\left[ \dot{z}_{\ell }+f(z_{\ell })\right] $, hence (\ref{conc})
is implied by%
\begin{eqnarray}
\frac{\dot{W}_{nm}}{W_{nm}}=-(M_{n}-M_{m})-\frac{\dot{z}_{n}-\dot{z}_{m}}{%
z_{n}-z_{m}} &&  \notag \\
+\sum_{\ell =1,\ell \neq n,m}^{N}\left[ \dot{z}_{\ell }+f(z_{\ell })\right]
\,\left( \frac{1}{z_{n}-z_{\ell }}-\frac{1}{z_{\ell }-z_{m}}\right) ~,\quad
n &\neq &m~.
\end{eqnarray}%
Inserting the logarithmic derivative of (\ref{BB1}) for $n\neq m$ and using (%
\ref{ddotzn}) we find by straightforward computations that the condition (%
\ref{conc}) is identically satisfied provided%
\begin{equation}
\dot{\varphi}_{n}=-M_{n}
\end{equation}%
and the functions $f$ and $\Phi $ satisfy the following functional
equations, 
\begin{equation}
f^{\prime }(x)+f^{\prime }(y)=2\frac{f(x)-f(y)}{x-y},\quad \Phi
(x)=f(x)f^{\prime }(x)
\end{equation}%
for all $x\neq y$. The general solution of these functional equations is
given by (\ref{f_Phi}), and this concludes our proof. Note that at the end
we can make the simplifying assignment $\varphi _{n}=M_{n}=0$.

\bigskip

\section*{IV. Alternative formulations}

The strategy to obtain alternative formulations of "goldfish-type"
$N$-body problems is by now standard (and quite old \cite{C1978}; for
a convenient up-to-date presentation see \cite{BC2006}). One
introduces a \textit{monic} polynomial $\psi (z,t)$ of degree $N$ in
$z,$ the $N$ zeros $z_{n}(t)$ of which evolve according to the
equations of motion of the $N$-body problem under consideration, and
then investigates the corresponding evolution of the $N$ coefficients
$c_{m}(t)$ of this polynomial. The route we follow to obtain the
equations of motions satisfied by the coefficients $c_{m}(t)$ --
equations that are of course no less \textit{solvable} than the
equations of motion satisfied by the zeros $z_{n}(t),$ since the
relationship among these quantities, the $N$ zeros $z_{n}$ and the $N$
coefficients $c_{m}$ of a polynomial of degree $N$, is purely
\textit{algebraic} -- is via the evolution equation satisfied by the
polynomial $\psi (z,t)$: note that this entails that this evolution
equation is itself \textit{solvable}. Since this technique is by now
standard, and appropriate formulas to implement it are available (see
in particular the Appendix in Ref. \cite{BC2006}), we present without
further ado the relevant results.

The evolution equation satisfied by the polynomial $\psi (z,t)$ (see (\ref%
{psi})) the zeros of which evolve according to the equations of motion (\ref%
{Gold}) reads%
\begin{eqnarray}
\psi _{tt}-2\,\left( z^{\,2}-a^{\,2}\right) \,\psi _{tz}+2\,\left[ \left(
N-2\right) \,z-c_{1}\right] \,\psi _{t} &&  \notag \\
+\left( z^{\,2}-a^{\,2}\right) ^{\,2}\,\psi _{zz}-2\,\left[ \left(
N-3\right) \,z-c_{1}\right] \,\left( z^{\,2}-a^{\,2}\right) \,\psi _{z} && 
\notag \\
+\left\{ N\,\left( N-5\right) \,z^{\,2}-2\,\left( N-2\right) \,c_{1}\,z+2\, 
\left[ 2\,N\,a^{\,2}+\dot{c}_{1}-c_{1}^{\,2}+3\,c_{2}\right] \right\} \,\psi
=0~. &&  \label{PDEGold}
\end{eqnarray}

\textit{Notation}: here and hereafter subscripted variables denote partial
differentiations with respect to them.

\textit{Remark 4.1}. This evolution equation, (\ref{PDEGold}), contains also
certain coefficients $c_{m}\equiv c_{m}(t),$ which are obviously (linearly)
related to the function $\psi (z,t),$ indeed clearly (see (\ref{psi}))%
\begin{equation}
c_{m}(t)=\left. \left[ \left( N-m\right) !\right] ^{-1}\,\frac{\partial
^{N-m}\,\psi (z,t)}{\partial \,z^{\,N-m}}\right\vert _{z=0}~.  \label{cpsi}
\end{equation}%
Hence (\ref{PDEGold}) is in fact a nonlinear functional equation satisfied
by the polynomial $\psi (z,t)$, and the fact that it is indeed satisfied by
a polynomial of degree $N$ in $z$, while not evident, is implied by the way
it has been obtained. $\boxdot $

From this evolution equation one obtains (using if need be the results in 
\cite{BC2006}) the corresponding system of ODEs satisfied by the
coefficients $c_{m}(t),$ see (\ref{AltGold}); and this of course justifies
the relevant results about the solvability of this system reported in the
preceding Section II.

Exactly the same procedure yields (\ref{AltIsoGold}) from (\ref{IsoGold}),
although a more direct route is via the "trick" formula (\ref{Trick}), as
indicated in the preceding Section II. Anyway we also display here, for
completeness, the equation for the polynomial $\tilde{\psi}(z,t)$ (see (\ref%
{Altpsi})) that provides the bridge connecting these two systems of ODEs
(deriving this equation is particularly easy using the formulas given in the
Appendix in Ref. \cite{BC2006}; but beware of the slight notational change
in the definition of the coefficients $c_{m}$ due to the $\left( i\right)
^{\,m}$ factor in the right-hand side of (\ref{Altpsi})):%
\begin{eqnarray}
\tilde{\psi}_{tt}-2\,z\,\left( z-i\right) \,\tilde{\psi}_{tz}+\,\left[
2\,\left( N-2\right) \,z-\left( 2\,N+1\right) \,i-2\,i\,\tilde{c}_{1}\right]
\,\tilde{\psi}_{t} &&  \notag \\
+z^{\,2}\,\left( z-i\right) ^{\,2}\,\tilde{\psi}_{zz}-2\,z\,\left(
z-i\right) \left[ N\,\left( z-i\right) -3\,z-i\,\tilde{c}_{1}\right] \,%
\tilde{\psi}_{z} &&  \notag \\
+\left[ N\,\left( N-5\right) \,z^{\,\,2}-2\,N^{\,2}\,i\,z-N\,(N+1)-2\,\left(
N-2\right) \,i\,\tilde{c}_{1}\,z-2\,\left( N-1\right) \,\tilde{c}_{1}\right.
&&  \notag \\
\left. +2\,(i\,\overset{\cdot }{\tilde{c}}_{1}+\tilde{c}_{1}^{\,2}-3\,\tilde{%
c}_{2})\right] \,\tilde{\psi}=0~. &&  \label{Isopsi}
\end{eqnarray}

\bigskip

\section*{V. Equilibrium configurations, behavior in their vicinity, \textit{%
Diophantine} relations}

In this section we discuss the equilibrium configurations (namely, the
time-independent solutions) of the \textit{isochronous }models (\ref{IsoGold}%
) and (\ref{AltIsoGold}) and the behavior of these models in the vicinity of
their equilibria. A motivation for focusing on the \textit{isochronous}
models is that they lead to the remarkable \textit{Diophantine} relations
reported at the end of Section II, as indicated below.

Clearly the equilibrium configuration $\tilde{c}_{m}(t)=\bar{c}_{m}$, $\dot{c%
}_{m}(t)=0$ of the system of ODEs (\ref{AltIsoGold}) is characterized by the
following system of $N$ algebraic equations:%
\begin{eqnarray}
&&-\left( m+2\right) \,\left( m-3\right) \,\bar{c}_{m+2}+2\,\left(
m-1\right) \,\left( m+1+\bar{c}_{1}\right) \,\bar{c}_{m+1}  \notag \\
&&+\left[ -m\,\left( m+1\right) -2\,(m-1)\,\bar{c}_{1}+2\,\bar{c}%
_{1}^{\,2}-6\,\bar{c}_{2}\right] \,\bar{c}_{m}=0~,  \notag \\
&&m=1,...,N~,~~~\bar{c}_{0}=1~,~~~\bar{c}_{-1}=\bar{c}_{N+1}=\bar{c}%
_{N+2}=0~.  \label{Eqcmbar}
\end{eqnarray}

Likewise the equilibrium configuration $\tilde{z}_{n}(t)=\bar{z}_{n}$, $%
\overset{\cdot }{\tilde{z}}_{n}=0,$ of the $N$-body problem (\ref{IsoGold})
is characterized by the following $N$ algebraic equations:%
\begin{equation}
\bar{z}_{n}\,(\bar{z}_{n}-i)\,\left[ \,\bar{z}_{n}+i+\sum_{m=1,m\neq n}^{N}%
\frac{\bar{z}_{m}\,(\bar{z}_{m}-i)\,}{\bar{z}_{n}-\bar{z}_{m}}\right] =0~.
\label{EqEq}
\end{equation}

These two configurations are related to each other by the polynomial formula
(see (\ref{Altpsi})) 
\begin{equation}
\bar{\psi}(z)=\dprod\limits_{n=1}^{N}\left( z-\bar{z}_{n}\right)
=\sum_{n=0}^{N}\left( i\right) ^{\,m}\,\bar{c}_{m}\,z^{\,N-m}~,~~\,\bar{c}%
_{0}=1~,  \label{psibar}
\end{equation}%
where $\bar{\psi}(z)$ is the "equilibrium" (namely, time-independent)
polynomial solution of (\ref{Isopsi}).

The general solution of the algebraic problem (\ref{EqEq}) can clearly be
broken down as follows: 
\begin{subequations}
\label{Break}
\begin{equation}
\bar{z}_{n}+i+\sum_{m=1,m\neq n}^{\nu }\frac{\bar{z}_{m}\,(\,\bar{z}%
_{m}^{{}}-i)\,}{\bar{z}_{n}-\bar{z}_{m}}=0~~~\text{for }n=1,...,\nu ~,
\label{Breaka}
\end{equation}%
\begin{equation}
\bar{z}_{n}=i~~~\text{for }n=\nu +1,...\mu ~,
\end{equation}%
\begin{equation}
\bar{z}_{n}=0~~~\text{for }n=\mu +1,...,N~,
\end{equation}%
with $\nu $ and $\mu $ \textit{nonnegative integers}, $0\leq \nu \leq \mu
\leq N$. \thinspace Of course each of these $3$ sectors will be empty if the
corresponding range of values of $n$ is empty (recall that $n=1,...,N$).

\textit{Remark 5.1}. In any equilibrium configuration the labeling of the
particles can be freely permuted. To write the breakdown (\ref{Break}) we
identified, without loss of generality, a (somewhat) definite assignment of
particle labels. $\boxdot $

\textit{Remark 5.2}. \textit{Genuine} equilibrium configurations of the $N$%
-body problem (\ref{IsoGold}) are characterized by the requirement that $%
\bar{z}_{n}\neq \bar{z}_{m}$ if $n\neq m$: indeed, whenever this condition
is violated, the equilibrium condition (\ref{EqEq}) becomes \textit{ambiguous%
} due to the vanishing of some denominator in the sum, compensated by a
vanishing of the corresponding numerator or by some other cancellation.
Hence a \textit{necessary} condition in order that the configuration
associated with the breakdown indicated in (\ref{Break}) correspond to a 
\textit{genuine} equilibrium configuration of the $N$-body problem (\ref%
{IsoGold}) is that $\mu \geq N-1$ and $\nu \geq \mu -1,$ so that at most one
of the $\bar{z}_{n}$'s vanishes (in which case we assign to it the highest
label, $\bar{z}_{N}=0$) and at most one takes the value $i$ (in which case
we assign to it the highest or next-to-highest label, $\bar{z}_{N}=i,$ or $%
\bar{z}_{N-1}=i$ if $\bar{z}_{N}=0$). But in the following it is convenient
to consider \textit{all} possible equilibrium configurations, including 
\textit{non genuine} ones, because, as we will see, such configurations,
while problematic to deal with in the context of the $N$-body problem (\ref%
{IsoGold}), correspond to equilibrium configurations $\bar{c}_{m}$ of the
system of ODEs (\ref{AltIsoGold}) which are instead perfectly legitimate in
the context of this \textit{nonlinear harmonic oscillators }model. Indeed
their consideration in such a context yields interesting findings (see
below). $\boxdot $

To get more information on the roots $\bar{z}_{n}$ of (\ref{Breaka}) we now
introduce a \textit{monic} polynomial $\varphi (z)$ of degree $\nu $ having
the $\nu $ numbers $\bar{z}_{n}$ with $n=1,...,\nu $ as its zeros: 
\end{subequations}
\begin{equation}
\varphi (z)=\dprod\limits_{n=1}^{\nu }\left( z-\bar{z}_{n}\right)
=\sum_{m=0}^{\nu }\left( i\right) ^{\,m}\,\varphi _{m}\,z^{\,\nu
-m}~,~~~\varphi _{0}=1~.  \label{phi}
\end{equation}%
Note that via this formula we also introduced the $\nu $ coefficients $%
\varphi _{m}$ of this polynomial.

It is now straightforward (and particularly easy using the formulas given in
the Appendix in Ref. \cite{BC2006}; but beware of the slight notational
change in the definition of the coefficients $\bar{c}_{m}$ due to the $%
\left( i\right) ^{\,m}$ factor in the right-hand side of (\ref{phi})) to
conclude that this polynomial must then satisfy the following equation,
implied by (\ref{Breaka}):%
\begin{eqnarray}
&&z^{\,2}\,\varphi ^{\prime \prime }-2\,\left( \nu -3\right) \,z\,\varphi
^{\prime }+\nu \,\left( \nu -5\right) \,\varphi  \notag \\
&=&i\,\left[ z\,\varphi ^{\prime \prime }-2\,\left( \nu +\varphi _{1}\right)
\,\varphi ^{\prime }\right] ~.  \label{Eqphi}
\end{eqnarray}

It is now easily seen that, via (\ref{phi}), this ODE (\ref{Eqphi}) yields
for the coefficients $\varphi _{m}$ the recurrence relation 
\begin{subequations}
\begin{equation}
m\,\left( m-5\right) \,\varphi _{m}=\left( m-\nu -1\right) \,\left( m+\nu
+2\,\varphi _{1}\right) \,\varphi _{m-1}~,  \label{Reca}
\end{equation}%
which must be complemented by the two extremal conditions (see (\ref{phi}))%
\begin{equation}
\varphi _{-1}=\varphi _{\nu +1}=0  \label{Recb}
\end{equation}%
and by the normalization condition (see (\ref{phi}))%
\begin{equation}
\varphi _{0}=1~.  \label{Recc}
\end{equation}

Clearly the two extremal conditions (\ref{Recb}) are identically satisfied
(for $m=0$ respectively $m=\nu +1$), while the condition (\ref{Recc}) yields
(for $m=1$) 
\end{subequations}
\begin{equation}
\varphi _{1}=\frac{\nu \,\left( \nu +1\right) }{2\,\left( 2-\nu \right) }~,
\label{c1bar}
\end{equation}%
entailing the requirement (hereafter assumed to hold)%
\begin{equation}
\nu \neq 2~.  \label{nunot2}
\end{equation}%
Insertion of (\ref{c1bar}) in (\ref{Reca}) yields finally the recursion 
\begin{equation}
m\,\left( m-5\right) \,\varphi _{m}=\left( m-\nu -1\right) \,\left( m+\frac{%
3\,\nu }{2-\nu }\right) \,\varphi _{m-1}~,
\end{equation}%
the solution of which is easily seen to exist only if $\nu \leq 5$. For the
remaining cases,%
\begin{equation}
\nu =0~\text{or~}1~\text{or~}3\text{~or~}4~\text{or }5~,  \label{nu1345}
\end{equation}%
see (\ref{nunot2}), we get the following solutions (recall (\ref{Recc})): 
\begin{subequations}
\label{phim}
\begin{equation}
\text{for }\nu =0~,~~\,\varphi (z)=\varphi _{0}=1~,  \label{phim0}
\end{equation}%
\begin{equation}
\text{for }\nu =1~,~~~\varphi _{0}=\varphi _{1}=1,~~~\varphi (z)=z+i~,
\label{phima}
\end{equation}%
\begin{equation}
\text{for }\nu =3~,~~~\varphi _{0}=1,~~\varphi _{1}=-6,~~\varphi
_{2}=14,~~\varphi _{3}=-14,~~,  \label{phimb}
\end{equation}%
\begin{equation}
\text{for~}\nu =4~,~~~\varphi _{m}=(-)^{\,m}\,\binom{5}{m}~,~~~m=0,1,...,4~,
\label{phime}
\end{equation}%
\begin{equation}
\text{for~}\nu =5~,~~~\varphi _{m}=(-)^{\,m}\,\binom{5}{m}%
~,~~~m=0,1,...,4~,~~~\varphi _{5}\text{ arbitrary~.}  \label{phimd}
\end{equation}

Via (\ref{Break}) and (\ref{phi}) it is clear that the \textit{monic}
polynomial $\bar{\psi}(z)$ of degree $N$ in $z$, see (\ref{psibar}) -- which
identifies as its $N$ zeros $\bar{z}_{n}$ respectively its $N$ coefficients $%
\bar{c}_{m}$ the equilibrium configurations of the models (\ref{IsoGold})
respectively (\ref{AltIsoGold}) -- is given by the formula 
\end{subequations}
\begin{equation}
\bar{\psi}(z)=\varphi (z)\,\left( z-i\right) ^{\,\mu -\nu }\,z^{\,N-\mu }~.
\label{psiphi}
\end{equation}%
It is thereby seen, via (\ref{phim}), that the coefficients $\bar{c}_{m}$
are given by the formulas (\ref{cmbar}) (with the \textit{arbitrary }%
constant $c=\varphi _{5}-1$ in (\ref{cmbarf}), see (\ref{phimd})).

\textit{Remark 5.3}. Clearly the coefficients $\bar{c}_{m}$ vanish for $%
m>\mu $, hence they \textit{all} vanish (except of course $\bar{c}_{0}=1$)
if $\mu =0$ (this assignment provides indeed a solution of (\ref{Eqcmbar})). 
$\boxdot $

Next, let us discuss the behavior of the system (\ref{AltIsoGold}) in the
neighborhood of its equilibrium configurations. To this end we set 
\begin{equation}
\tilde{c}_{n}(t)=\bar{c}_{n}+\varepsilon \,\rho _{n}(t)+O(\varepsilon
^{\,2})~,  \label{cepsi}
\end{equation}%
with $\bar{c}_{m}$ the coefficients $c_{m}$ at equilibrium (as determined
above) and $\varepsilon $ a small parameter. We thereby obtain in the
standard manner the \textit{linearized} equations of motion%
\begin{eqnarray}
\ddot{\rho}_{m}+2\,\left( m-1\right) \,i\,\dot{\rho}_{m+1}-\left( 2\,m+1+2\,%
\bar{c}_{1}\right) \,i\,\dot{\rho}_{m} &&  \notag \\
-\left( m+2\right) \,\left( m-3\right) \,\rho _{m+2}+2\,\left( m-1\right)
\,\left( m+1+\bar{c}_{1}\right) \,\rho _{m+1} &&  \notag \\
+\left[ -m\,\left( m+1\right) -2\,(m-1)\,\bar{c}_{1}+2\,\bar{c}_{1}^{\,2}-6\,%
\bar{c}_{2}\right] \,\rho _{m} &&  \notag \\
+2\,i\,\bar{c}_{m}\,\dot{\rho}_{1}+2\,\left[ \left( m-1\right) \,\bar{c}%
_{m+1}-(m-1-2\,\bar{c}_{1})\,\bar{c}_{m}\right] \,\rho _{1}-6\,\bar{c}%
_{m}\,\rho _{2}=0~, &&  \notag \\
m=1,...,N~,~~~\rho _{0}=0~,~~~\rho _{N+1}=\rho _{N+2}=0~. &&  \label{Eqgamma}
\end{eqnarray}

The general solution of this linear system of ODEs, (\ref{Eqgamma}), reads 
\begin{equation}
\rho _{m}(t)=\sum_{n=1}^{N}\left[ a_{n}^{(+)}\,r_{m}^{(+)(n)}\,\exp \left(
i\,p_{n}^{(+)}\,t\right) +a_{n}^{(-)}\,r_{m}^{(-)(n)}\,\exp \left(
i\,p_{n}^{(-)}\,t\right) \right] ~,  \label{Solgamma}
\end{equation}%
where the $2\,N$ numbers $a_{n}^{(\pm )}$ are \textit{arbitrary} (to be
fixed by the initial data) while the $2\,N$ numbers $p_{n}^{(\pm )}$,
respectively the $2\,N$ corresponding ($t$-independent) $N$-vectors $%
\underline{r}^{(\pm )(n)}\equiv \left( r_{1}^{(\pm )(n)},...,r_{N}^{(\pm
)(n)}\right) ,$ are the eigenvalues, respectively the eigenvectors, of the ($%
N$-vector) generalized eigenvalue equation (\ref{Eigena}). This implies (\ref%
{Eigenb}) with the two $N\times N$ matrices $A$ and $B$ defined
(componentwise) by the formulas (\ref{AB}). But we know (see \textbf{%
Proposition 2.13}) that \textit{all} the \textit{nonsingular} solutions of
the system of \textit{nonlinear harmonic oscillators} (\ref{AltIsoGold}) are 
\textit{completely periodic} with period $2$\thinspace $\pi $, hence the
(certainly \textit{nonsingular}) solutions describing the behavior of this
system around equilibrium must have the same periodicity property, implying
that \textit{all} the eigenvalues $p_{n}^{\left( \pm \right) }$ yielded by
the generalized eigenvalue problem (\ref{Eigena}) must be \textit{integers.}
And this entails the validity of \textbf{Proposition 2.14}.

\textit{Remark 5.4}. In the special case of the equilibrium configuration $%
\bar{c}_{m}=0$ (for $m=1,...,N$, while of course $\bar{c}_{0}=1;$ see the 
\textit{Remark 5.3}), the matrices $A$ and $B$ become \textit{triangular}
and the computation of the eigenvalues $p_{n}^{\left( \pm \right) }$ is then
a trivial task, yielding 
\begin{equation}
p_{n}^{\left( +\right) }=n+1~,~~~p_{n}^{\left( -\right) }=n~.~\boxdot
\label{Diopha}
\end{equation}

\bigskip

\section*{VI. Outlook}

It is remarkable that a research project started with the main purpose to
clarify a methodological issue -- namely, the relationship among two
different approaches to the same question: that of identifying \textit{%
solvable} many-body problems -- resulted in the identification of a \textit{%
novel} solvable many-body problem. To the readers who might imagine -- in
view of the recent discovery of several such new models, as reviewed in Ref. 
\cite{BC2006} -- that this is a relatively trivial task, we suggest to try
and find themselves some new model. Our educated guess is that such a task
is quite challenging. We are nevertheless ourselves hopeful that new
many-body models exist, and that they might be discovered/manufactured by
the techniques described in this paper. In any case this possibility remains
as a tantalizing prospect, until a way is found to ascertain conclusively
that these approaches have exhausted their capability to yield many-body
models of the kind investigated herein which are both new and interesting
(although the second of these two qualities involves of course a value
judgment).

Another research direction (perhaps suitable as a PhD project) is towards
proving the \textit{Diophantine} conjectures proffered in this paper (see
the end of Section II) and in previous ones (see \cite{BC2006} and other
papers referred to there), as well as obtaining additional findings of this
kind (for instance by applying techniques analogous to those of Section V to
the model (\ref{AltGold}), taking advantage of the results reported in the
Appendix).

\bigskip

\section*{Acknowledgments}

We gratefully acknowledge financial support by the European Union project
RTN\ "Enigma" and the "G\"{o}ran Gustafsson Foundation" which made possible
a two-week visit to Sweden in February 2006 by one of us (FC) when the
results reported in this paper emerged. One of us (FC) would also like to
thank the Japan Society for the Promotion of Science (JSPS) and the Yukawa
Institute at Kyoto University, in particular Ryu Sasaki, for the pleasant
hospitality during a six-week visit to Japan in March-April 2006, when this
work was (almost) completed. It is moreover a pleasant duty to acknowledge
with thanks the assistance provided to one of us (FC) by Ryu Sasaki by
performing some computer-aided checks, by Andrea La Malfa with the handling
of Mathematica and by V. I. Inozemtsev with useful suggestions concerning
references. One of us (EL) is supported by the Swedish Science Research
Council (VR).

\bigskip
\app

\section*{Appendix: Equilibrium configurations of the models (\protect\ref%
{AltGold}) and (\protect\ref{Gold})}

In \textit{Remark 2.9 }it was mentioned that there exist additional
equilibrium (i. e., time-independent) solutions of the model (\ref{AltGold})
besides (\ref{EquiGoldSpec}). In this appendix we firstly list \textit{all}
these equilibrium configurations (as obtained via Maple) for $N=2,3,4$ and $%
5 $. We then outline a technique allowing to obtain \textit{all} the
equilibrium configurations for \textit{arbitrary} $N$ (we of course did
check that these findings reproduce, for $N=2,3,4$,$5$, those obtained via
Maple).

These equilibrium configurations are clearly solutions of the following set
of $N$ algebraic equations (see (\ref{AltGold})):

\begin{eqnarray}
&&\left( m+2\right) \,\left( m-3\right) \,c_{m+2}-2\,\left( m-1\right)
\,c_{1}\,c_{m+1}  \notag \\
&&+2\,\left[ m\,\left( N+2-m\right) \,a^{\,2}-c_{1}^{\,2}+3\,c_{2}\right]
\,c_{m}  \notag \\
&&-2\,\left( N+1-m\right) \,a^{\,2}\,c_{1}\,c_{m-1}+\left( N+2-m\right)
\,\left( N+1-m\right) \,a^{\,4}\,c_{m-2}=0~,  \notag \\
m &=&1,...,N~,~~~c_{0}=1~,~~~c_{-1}=c_{N+1}=c_{N+2}=0~.  \label{A1}
\end{eqnarray}

The following solutions of this algebraic system have been obtained via
Maple.

For $N=2$ 
\begin{subequations}
\label{EquiGoldGen}
\begin{equation}
c_{1}=0~,~~~c_{2}=-a^{\,2}~,  \label{EquiGoldGena}
\end{equation}%
or%
\begin{equation}
c_{2}=\frac{c_{1}^{\,2}}{3}-\frac{a^{\,2}}{3}~,~~~c_{1}\text{ \ \ arbitrary}
\label{EquiGoldGenb}
\end{equation}%
(the equilibrium configurations (\ref{EquiGoldSpec}) correspond to the
latter one, (\ref{EquiGoldGenb}), with $c_{1}=\pm 2\,a)$.

For $N=3$, 
\end{subequations}
\begin{equation}
c_{2}=\frac{c_{1}^{\,2}}{3}\pm \frac{a\,c_{1}}{3}-a^{\,2}~,~~~c_{3}=\pm 
\frac{a\,c_{1}^{\,2}}{3}-\frac{2\,a^{\,2}\,c_{1}}{3}~,~~\,c_{1}\text{
\thinspace \thinspace arbitrary}
\end{equation}%
(the equilibrium configurations (\ref{EquiGoldSpec}) obtain for $c_{1}=\pm
3\,a)$.

For $N=4$, 
\begin{subequations}
\begin{equation}
c_{2}=\frac{c_{1}^{\,2}}{3}-\frac{4\,a^{\,2}}{3}~,~~~c_{3}=-a^{\,2}%
\,c_{1}~,~~\,c_{4}=-\frac{a^{\,2}\,c_{1}^{\,2}}{3}+\frac{a^{\,4}}{3}%
~,~~\,c_{1}\text{ \thinspace \thinspace arbitrary~,}
\end{equation}%
or%
\begin{eqnarray}
c_{2} &=&\frac{c_{1}^{\,2}}{3}\pm \frac{2\,a\,c_{1}}{3}-2\,a^{%
\,2}~,~~~c_{3}=\pm \frac{2\,a\,c_{1}^{\,2}}{3}-\frac{5\,a^{\,2}\,c_{1}}{3}~,
\notag \\
c_{4} &=&\frac{a^{\,2}\,c_{1}^{\,2}}{3}\mp \frac{4\,a^{\,3}\,c_{1}}{3}%
+a^{\,4}~,~~~c_{1}\text{ \thinspace \thinspace arbitrary}  \label{cN4}
\end{eqnarray}%
(the equilibrium configurations (\ref{EquiGoldSpec}) obtain from (\ref{cN4})
for $c_{1}=\pm 4\,a)$.

For $N=5$, 
\end{subequations}
\begin{subequations}
\begin{eqnarray}
c_{2} &=&\frac{c_{1}^{\,2}}{3}\pm \frac{a\,c_{1}}{3}-2\,a^{\,2}~,~~~c_{3}=%
\pm \frac{a\,c_{1}^{\,2}}{3}-\frac{5\,a^{\,2}\,c_{1}}{3}~,~~\,c_{4}=-\frac{%
a^{\,2}\,c_{1}^{\,2}}{3}\mp \frac{a^{\,3}\,c_{1}}{3}+a^{\,4}~,  \notag \\
c_{5} &=&\mp \frac{a^{\,3}\,c_{1}^{\,2}}{3}+\frac{2\,a^{\,4}\,c_{1}}{3}%
~,~~~\,c_{1}\text{ \thinspace \thinspace arbitrary~,}
\end{eqnarray}%
or%
\begin{eqnarray}
c_{2} &=&\frac{c_{1}^{\,2}}{3}\pm a\,c_{1}-\frac{10\,\,a^{\,2}}{3}%
~,~~~c_{3}=\pm
a\,c_{1}^{\,2}-3\,a^{\,2}\,c_{1}~,~~~c_{4}=a^{\,2}\,c_{1}^{\,2}\mp
5\,a^{\,3}\,c_{1}+5\,a^{\,4}~,  \notag \\
c_{5} &=&\pm \frac{a^{\,3}\,c_{1}^{\,2}}{3}-2\,a^{\,4}\,c_{1}\pm \frac{%
8\,a^{\,5}}{3}~,~~~c_{1}\text{ \thinspace \thinspace arbitrary}  \label{cN5}
\end{eqnarray}%
(the equilibrium configurations (\ref{EquiGoldSpec}) obtain from (\ref{cN5})
for $c_{1}=\pm 5\,a)$.

The route we follow to obtain \textit{all }the solutions of the system (\ref%
{A1}) is analogous to that followed in Section V. The starting point is to
introduce a \textit{monic} polynomial $\bar{\psi}(z)$ of degree $N$ that has
the $N$ numbers $c_{m}$ solutions of (\ref{A1}) as its $N$ coefficients: 
\end{subequations}
\begin{equation}
\bar{\psi}(z)=\dprod\limits_{n=1}^{N}\left( z-\bar{z}_{n}\right)
=\dsum\limits_{m=0}^{N}c_{m}\,z^{\,N-m}~,~~~c_{0}=1~.  \label{A2}
\end{equation}%
Note the analogy of these formulas with (\ref{psi}), and the fact that we
also introduced the $N$ zeros $\bar{z}_{n}$ of this polynomial $\bar{\psi}%
(z),$ which clearly provide the equilibrium configuration of the $N$-body
problem (\ref{Gold}) (although not necessarily a \textit{genuine}
equilibrium configuration), hence satisfy the following system of $N$
algebraic ODEs:%
\begin{equation}
(\bar{z}_{n}^{\,2}-a^{\,2})\,\left[ \bar{z}_{n}+\sum_{m=1,m\neq n}^{N}\frac{%
\bar{z}_{m}^{\,2}-a^{\,2}}{\bar{z}_{n}-\bar{z}_{m}}\right] =0~.  \label{A3}
\end{equation}

Our strategy to find all the solutions of the system (\ref{A1}) is to find
firstly all the solutions of this system, (\ref{A3}), and then use (\ref{A2}%
).

Clearly the solutions of (\ref{A3}) can be broken down as follows: 
\begin{subequations}
\label{A4}
\begin{equation}
\bar{z}_{n}+\sum_{m=1,m\neq n}^{\nu }\frac{\bar{z}_{m}^{\,2}-a^{\,2}}{\bar{z}%
_{n}-\bar{z}_{m}}=0~~~\text{for}~~~n=1,...,\nu ~,  \label{A4c}
\end{equation}%
\begin{equation}
\bar{z}_{n}=a~~~\text{for}~~~n=\nu +1,...,\nu +\mu ~,  \label{A4a}
\end{equation}%
\begin{equation}
\bar{z}_{n}=-a~~~\text{for}~~~n=\nu +\mu +1,...,N~,  \label{A4b}
\end{equation}%
with the two \textit{nonnegative integers} $\nu $ and $\mu $ \textit{%
arbitrary} except for the constraint%
\begin{equation}
\nu +\mu \leq N~  \label{A4d}
\end{equation}%
(implying of course that neither one of these two nonnegative integers can
exceed $N$).

This assignment, (\ref{A4}), clearly entails that 
\end{subequations}
\begin{subequations}
\label{A5}
\begin{equation}
\bar{\psi}(z)=\left( z-a\right) ^{\,\mu }\,\left( z+a\right) ^{\,N-\mu -\nu
}\,\phi _{\nu }(z)~,  \label{A5a}
\end{equation}%
with $\phi _{\nu }(z)$ the \textit{monic} polynomial of degree $\nu ,$%
\begin{equation}
\phi _{\nu }(z)=\dprod\limits_{n=1}^{\nu }\left( z-\bar{z}_{n}\right)
=\dsum\limits_{m=0}^{\nu }f_{m}\,z^{\,\nu -m}~,~~~f_{0}=1~,  \label{A5b}
\end{equation}%
the zeros of which satisfy the algebraic relations (\ref{A4c}). Hence this
polynomial $\phi _{\nu }(z)$ satisfies the equation 
\end{subequations}
\begin{equation}
\left( z^{\,2}-a^{\,2}\right) \,\phi _{\nu }^{\prime \prime }-2\,\left[
\left( \nu -3\right) \,z-f_{1}\right] \,\phi _{\nu }^{\prime }+\nu \,\left(
\nu -5\right) \,\phi _{\nu }=0~,  \label{A6}
\end{equation}%
as implied by the (by now standard) technique to transform algebraic
equations such as (\ref{A4c}) into differential equations (see for instance
the Appendix in Ref. \cite{BC2006}). Here and below primes denote of course
differentiations with respect to the argument of the function they are
appended to.

Before proceeding to discuss the solution of this equation let us consider
the special case with $\nu =0$ entailing $\phi _{0}(z)=1$ (which solves (\ref%
{A6}) trivially). In this case (\ref{A5a}) and (\ref{A4d}) yield 
\begin{subequations}
\begin{equation}
\bar{\psi}(z)=\left( z-a\right) ^{\,\mu }\,\left( z+a\right) ^{\,N-\mu
}~,~~~\mu =0,1,...,N~.  \label{A8a}
\end{equation}%
It is then easily seen from (\ref{A2}) that this entails%
\begin{equation}
c_{m}=a^{\,m}\,\sum_{\ell =\max (0,m+\mu -N)}^{\min (\mu ,m)}\left( -\right)
^{\,\ell }\,\binom{\mu }{\ell }\,\binom{N-\mu }{m-\ell }~.  \label{A8b}
\end{equation}%
This formula provides a set of equilibrium configurations, characterized by
the integer $\mu $ in the range $0\leq \mu \leq N$; in particular the two
solutions corresponding to $\mu =0$ and to $\mu =N$ are easily seen to yield
the two solutions (\ref{EquiGoldSpec}).

Let us now return to (\ref{A6}), assuming hereafter that the integer $\nu $
is \textit{positive}, $\nu >0$ (to avoid unnecessary notational
complications). To solve this equation, (\ref{A6}), we set 
\end{subequations}
\begin{equation}
\phi _{\nu }(z)=a^{\,\nu }\,\chi (x)~,~\,\,z=a\,\left( x-1\right) ~.
\label{A9}
\end{equation}%
This formula implies that $\chi (x)$ is again a \textit{monic }polynomial of
degree $\nu $ (although for notational simplicity we do not signal this via
a subscript $\nu $). We also set (in analogy to (\ref{A5b}))%
\begin{equation}
\chi (x)=\dprod\limits_{n=1}^{\nu }\left( x-x_{n}\right)
=\dsum\limits_{m=0}^{\nu }\chi _{m}\,x^{\,\nu -m}~,~~~\chi _{0}=1~,
\label{A10}
\end{equation}%
and we then note that this formula, together with (\ref{A5b}), entails 
\begin{subequations}
\label{A12}
\begin{equation}
f_{m}=a^{\,m}\,\sum_{\ell =0}^{m}\binom{\nu -\ell }{m-\ell }\,\chi _{\ell }~,
\label{A12a}
\end{equation}%
hence in particular%
\begin{equation}
f_{1}=a\,\left( \nu +\chi _{1}\right) ~.  \label{A12b}
\end{equation}

Via this formula and (\ref{A9}) the differential equation (\ref{A6}) now
reads 
\end{subequations}
\begin{subequations}
\begin{equation}
x\,\left( x-2\right) \,\chi ^{\prime \prime }-2\,\left[ \left( \nu -3\right)
\,x-2\,\nu +3-\chi _{1}\right] \,\chi ^{\prime }+\nu \,\left( \nu -5\right)
\,\chi =0~,  \label{A13a}
\end{equation}%
entailing, via (\ref{A10}), the two-term recurrence%
\begin{equation}
m\,\left( m-5\right) \,\chi _{m}=2\,\left( \nu +1-m\right) \,\left( 3-\nu
-\chi _{1}-m\right) \,\chi _{m-1}~,  \label{A13b}
\end{equation}%
implying (for $m=0$ and $m=\nu +1$) the extremal conditions $\chi _{-1}=\chi
_{\nu +1}=0$ (consistently with (\ref{A10})).

For $m=1$ the recurrence formula (\ref{A13b}) (together with $\chi _{0}=1$,
see (\ref{A10})) yields the relation 
\end{subequations}
\begin{equation}
\left( \nu -2\right) \,\chi _{1}=-\nu \,\left( \nu -2\right) ~,  \label{A14}
\end{equation}%
requiring that the two cases with $\nu =2$ and $\nu \neq 2$ be treated
separately.

For $\nu =2$ one easily obtains the solution 
\begin{subequations}
\begin{equation}
\chi _{0}=1~,~\,~\,\chi _{1}~\text{arbitrary,}~~\,\chi _{2}=\frac{\chi
_{1}\,\left( \chi _{1}+1\right) }{3}~.  \label{A15a}
\end{equation}

\textit{Remark A.1}. Let us note as a curiosity that the recursion (\ref%
{A13b}) with $\nu =2$ allows this solution (\ref{A15a}) to be extended as
follows:%
\begin{eqnarray}
\chi _{3} &=&\chi _{4}=0~,~~~\chi _{5}~\text{also arbitrary,}  \notag \\
\chi _{m} &=&\frac{2^{\,m-3}\,15}{m\,\left( m-1\right) \,\left( m-2\right) }%
\,\binom{m-1+\chi _{1}}{m-5}\,\chi _{5}~,~~~m=5,6,...~.
\end{eqnarray}%
But we are only interested in the solution (\ref{A15a}) with $\chi _{m}=0$
for $m>2,$ entailing that $\chi (x)$ is a polynomial of degree $\nu
=2~.~\boxdot $

So in this $\nu =2$ case we get 
\end{subequations}
\begin{equation}
\chi (x)=x^{2}+\chi _{1}\,x+\frac{\chi _{1}\,\left( \chi _{1}+1\right) }{3}
\end{equation}%
hence, via (\ref{A9}), 
\begin{subequations}
\begin{equation}
\phi _{2}(z)=z^{\,2}+\,f_{1}\,z+\frac{f_{1}^{\,2}-a^{\,2}}{3}~,~~~f_{1}~%
\text{arbitrary}
\end{equation}%
where we set, consistently with (\ref{A5b}), $f_{1}=\left( 2+\chi
_{1}\right) \,a$.

For $\nu \neq 2$ the recursion (\ref{A13b}) with (\ref{A14}) yields 
\end{subequations}
\begin{eqnarray}
\chi _{0} &=&1~,~~\,\,\chi _{1}=-\nu ~,~~\,\,\chi _{2}=\frac{\nu \,\left(
\nu -1\right) }{3}~,~~~\chi _{3}=\chi _{4}=0~,~~~\chi _{5}\text{ arbitrary,}
\notag \\
\chi _{m} &=&\frac{\left( -\right) ^{\,m-1}\,2^{\,m-3}\,15}{m\,\left(
m-1\right) \,\left( m-2\right) }\,\binom{\nu -5}{m-5}\,\chi
_{5}~,~~~\,m=6,...,\nu ~.  \label{A16}
\end{eqnarray}%
Of course the second line of this equation is only relevant if $\nu >5,$
which can only happen if $N>5.$

From these results, via (\ref{A5a}) with (\ref{A4d}) and (\ref{A9}), we
arrive finally, after a bit of trivial algebra, at the following two
determinations of the polynomial $\bar{\psi}(z)$ (see (\ref{A2})): 
\begin{subequations}
\label{A17}
\begin{equation}
\bar{\psi}(z)=\left( z^{\,2}+c\,z+\frac{c^{\,2}-a^{\,2}}{3}\right) \,\left(
z-a\right) ^{\,\mu }\,\left( z+a\right) ^{\,N-2-\mu }\,~,~~~\mu
=0,1,...,N-2~,  \label{A17b}
\end{equation}%
\begin{eqnarray}
\bar{\psi}(z)=\left( z-a\right) ^{\,\mu }\,\left( z+a\right) ^{\,N-\mu }\,%
\left[ 1-\frac{\nu \,a}{z+a}+\frac{\nu \,\left( \nu -1\right) \,a^{\,2}}{%
3\,\left( z+a\right) ^{\,2}}\right. &&  \notag \\
\left. +c\,\sum_{\ell =0}^{\nu -5}\frac{\left( -2\right) ^{\,\ell }}{\left(
\ell +5\right) \,\left( \ell +4\right) \,\left( \ell +3\right) }\,\binom{\nu
-5}{\ell }\,\left( \frac{a}{z+a}\right) ^{\,\ell +5}\right] ~, &&  \notag \\
\nu =5,6,...,N~,~~~\mu =0,1,...,N-\nu ~. &&  \label{A17e}
\end{eqnarray}%
The first, (\ref{A17b}), of these two formulas is applicable for $N\geq 2$,
having been obtained from the previous results corresponding to $\nu =2,$
with $c=f_{1}$ an \textit{arbitrary }number. It includes the results
obtainable from the cases with $\nu =1$ and $\nu =3$; likewise, the result
corresponding to $\nu =4$ has not be reported, as it is encompassed by the
result (\ref{A8a}). The second, (\ref{A17e}), of these two formulas is of
course only applicable provided $N\geq 5$, having being obtained from (\ref%
{A16}) for $\nu \geq 5,$ with $c=\chi _{5}\,/\,60$ an \textit{arbitrary}
number. Together with (\ref{A8a}) these two formulas determine \textit{all}
the equilibrium configurations $c_{m}$ of the system (\ref{AltGold}) (and as
well \textit{all} the equilibrium configurations -- not necessarily \textit{%
genuine} -- of the $N$-body problem (\ref{Gold})) -- up to a final (trivial
but tedious) step, to be performed using (\ref{A2}) (as done above to obtain
(\ref{A8b}) from (\ref{A8a})), which we leave as a task for the diligent
reader.

\appende

\end{subequations}

\end{document}